\documentstyle[aas2pp4]{article}

\newcommand{\dsigw}{\mbox{$\dot\Sigma_{\rm w}$}{}}
\newcommand{\D}{\partial}
\newcommand{\dotl}{\mbox{$\dot{\it l}$}{}}
\newcommand{\nuw}{\mbox{$\nu_{\rm w}$}{}}

\begin{document}
\title{Accretion Disk Evolution With Wind Infall I.  General
Solution and Application to Sgr~A*}
\author{Heino Falcke}
\affil{Astronomy Department, University of Maryland, College Park,
MD 20742-2421 (hfalcke@astro.umd.edu)}
\author{Fulvio Melia\altaffilmark{1}}
\affil{Department of Physics \& Steward Observatory, University of
Arizona, Tucson, AZ 85721 (melia@as.arizona.edu)}
\altaffiltext{1}{Presidential Young Investigator}

\begin{abstract}
The evolution of an accretion disk can be influenced significantly
by the deposition of mass and angular momentum by an infalling
Bondi-Hoyle wind.  Such a mass influx impacts the long-term
behavior of the disk by providing additional sources of viscosity
and heating.  In this paper, we derive and solve the disk
equations when these effects are taken into account.  We present
a survey of models with various wind configurations and
demonstrate that the disk spectrum may then differ substantially
from that of a standard $\alpha$-disk.  In particular, it is
likely that a wind-fed disk has a significant infrared bump due
to the deposition of energy in its outer region.  We apply some
of the results of our calculations to the Galactic Center black
hole candidate Sgr~A* and show that if a fossil disk is present
in this source, it must have a very low viscosity parameter ($\alpha \la
10^{-4}$) and the Bondi-Hoyle wind must be accreting with a very
high specific angular momentum to prevent it from circularizing
in the inner disk region where its impact would be most noticeable.
\end{abstract}


\section{Introduction}
\subsection{Disk vs. Spherical Accretion}
Compact objects, such as white dwarfs, neutron stars, and black holes
may accrete either from the ambient medium (generally the Interstellar
Medium --- ISM) or from
a companion in a binary configuration.  Whether or not this accretion
is mediated via a disk or it constitutes a more or less quasi-spherical
infall depends critically on the specific angular momentum carried by
the captured gas. It therefore is not always clear whether disk or
quasi-spherical accretion dominates or if both are present.

For example, in Roche Lobe overflow, the gas circularizes quickly 
and forms a disk that can extend as far out as the binary orbital 
midpoint (e.g., Lamb \& Melia 1987; Melia \& Lamb 1987).  
Its temperature is generally too low to provide
any significant support in the vertical direction, and so the disk is
relatively flat and optically thick, at least until the gas approaches
the inner regions close to the accreting object. Some binary systems
are too wide for the companion to fill its Roche Lobe, and for these
the accretion proceeds primarily via the capture of a portion of its
wind by the compact object.  In many circumstances, the latter's magnetic 
field is quite strong and the plasma is entrained onto its polar cap regions
(e.g., Burnard, Klein \& Arons 1988).  Otherwise, the gas can either
form a small (highly variable) disk whose size is specified by the
circularization radius dictated by the specific angular momentum, or
it may fall directly onto the accretor (Taam \& Fryxell 1988).

In active galactic nuclei (AGN), the central, massive black hole
generally accretes from the ambient medium. This can either be the ISM gas
surrounding the nucleus or it may be the winds produced by stars in
the central cluster. Although it is likely that Bondi-Hoyle 
accretion dominates the gaseous motions at large radii (out at the
so-called accretion radius of order $0.1-1$ pc), there is ample evidence
that much of the activity in AGNs is associated with an accretion
disk at smaller radii.  It appears, therefore, that the quasi-spherical
accretion at large radii gradually settles into a planar, more or less
Keplerian structure toward the central accretor, i.e., the accretion
disk in these sources is `fed' by the quasi-spherical infall.

\subsection{Scope of the Paper}
In this paper, we begin our study of the effects on the accretion disk
evolution due to the continued infall from quasi-spherical accretion.
We rederive the disk equations taking into account the contribution
from the wind to the overall disk dissipation, the disk temperature,
its column density, the accretion rate through the disk and the
overall disk spectrum.  All of these are expected to differ
substantially between the isolated systems fed within a plane from
their outer radii and those in which a disk is mainly fed from above.

The infalling gas can differ from the disk plasma in several
respects: (i) its specific angular momentum at a given radius may
be close to Keplerian (like that of the disk), or it may be much 
less, perhaps even zero, (ii) its mass flux may be much larger than 
that through the disk, and (iii) its ram pressure may differ 
significantly from that of the disk material, and thereby produce 
its own dynamical influence on the gas closer to the compact object.

We wish to study the impact of such a mass influx on the long-term
behavior of the disk. In future work, we will couple this analysis
with 3D hydrodynamic (and possibly magnetohydrodynamic) simulations of
the ambient infall.  This will provide much more realistic profiles of
accreted mass and specific angular momentum as functions of radius in
the disk. The process we discuss may be relevant to galactic nuclei as
well as stellar mass black holes, neutron stars and star
formation. Here we will concentrate on the former.

\subsection{The Galactic Center}
We shall specifically consider the example of the Galactic Center, where the
presence of a supermassive black hole (coincident with the unique
radio source Sgr~A*) and strong, nearby stellar winds should lead to 
substantial Bondi-Hoyle accretion (e.g., Melia 1992; Ozernoy 1993;  
Ruffert \& Melia 1994). In addition, stars may get tidally disrupted as 
they approach the strong central gravitational field.  Depending on the
strength of the disruption, as much as half of the stellar mass may be
left as a remnant surrounding the black hole (Khohklov \& Melia 1995
and references cited therein, particularly Rees 1990 and Shlossman, Begelman
\& Frank 1990).  This remnant may, under some circumstances, form a relatively
cold disk that evolves without a substantial mass infusion at its
outer edge. Otherwise, there might also be a fossil accretion disk
around the central black hole which is the remnant of a phase of higher
activity in the past (Falcke \& Heinrich 1994). In both cases the disk
would inevitably interact with the infalling wind and thus change the
matter inflow and the structure of the accretion disk. The resulting
observational signature will then depend strongly on the ratio of wind 
to disk accretion rates, the viscosity in the disk, the specific angular 
momentum of the infalling matter and the time scales involved.

The broad-band radiative emission from Sgr~A* may be produced either in
the quasi-spherical accretion portion of the inflow (Melia 1992 \& 1994) 
with $\dot M_{\rm w}\sim 10^{21-22}$ g s$^{-1}$, or by a combination
of disk plus radio-jet (Falcke et al. 1993a\&b; see Falcke 1996 for 
a review). As discussed above, quasi-spherical accretion seems to be 
unavoidable at large radii, but the low actual luminosity of Sgr~A* 
seems to point to a much lower accretion rate in a starving disk and 
hence Sgr~A* appears to be the ideal case for applying the concept 
of a fossil disk fed by quasi-spherical accretion.  We note here that
the disk may also be advective, in which case the luminosity may be 
decreased substantially due to the large fraction of dissipated energy 
carried inwards through the event horizon by the disk plasma (e.g., 
Narayan et al. 1995).  In future developments of our model, we shall
also consider such disk configurations.

The outline of the present paper is as follows: in Sec.~2 we 
discuss the wind energy deposition onto the disk and we will derive the
equations for the evolution of an $\alpha$-disk with arbitrary wind
infall in Sec.~3; in Sec.~4 we will present numerical solutions for various
parameters and discuss their implication for the Galactic Center in
Section 5. The paper is summarized in Sec.~6, where we also
describe further applications of our model.

\section{Wind Energy Deposition Onto The Disk}
When deriving the modified disk equations below, we will need to
understand, at least crudely, the effects of the impacting wind on the
disk, and where the energy is deposited.  Since the inflow is usually
highly supersonic near the surface of the disk (e.g., Coker et
al. 1996, Paper II), it is reasonable to assume that the wind
termination proceeds via a shock.  To estimate the relevant time
scales, we can make the rough approximation that an axisymmetric wind
approaches the flat disk (of radius $R_{\rm d}=10^{16}\, {\rm cm}\cdot
R_{16}$) in free fall, with an accretion rate $\dot M_{\rm w}
=10^{-4}M_\odot\cdot\dot M_{-4}$.  The black hole mass is taken to be
$M_{\bullet}=10^6\,M_\odot\cdot M_{6}$.  As we shall see,
these parameters are typical for weakly accreting nuclei, such as Sgr
A* at the Galactic Center.  For a strong shock, the downstream flow
has a velocity
\begin{equation}
v_{\rm sh}\approx{1\over 4}\sqrt{2 G M_\bullet\over R}\approx 4\times 10^7\,{\rm cm\over
sec}\;\sqrt{M_{6}\over R_{16}}
\end{equation}
and a density 
\begin{equation}
n_{\rm sh}\approx\dot {M_{\rm w} \over \pi R^2 v_{\rm sh} m_{\rm p}}\approx
3\times 10^5\,{\rm cm}^{-3}\;{\dot M_{-4}\over\sqrt{M_{6} R_{16}^3}},
\end{equation}
while the temperature of the shocked gas is given by
\begin{equation}
T_{\rm sh}={3 m_{\rm p} v_{\rm sh}^2\over k_{\rm B}}=6\times 10^7\,{\rm
K}\; {M_{6}\over R_{16}}.
\end{equation}

Behind the shock, the gas will have to settle onto the disk, which in
our case is much cooler with a temperature $T_{\rm d}$ of several 100
to 1000 K.  In this situation, the dominant radiative cooling
mechanism of the hot shocked gas with energy density $dE/dV=3/2 k_{\rm
B} T_{\rm sh} n_{\rm sh}$ would be thermal Bremsstrahlung
($\epsilon_{\rm brems}=1.7\times 10^{-27}\;[T/{\rm K}]^{1/2} [n/{\rm
cm}^{-3}]^2\,{\rm erg\,sec^{-1} cm^{-3}}$), corresponding to a cooling
time scale of $t_{\rm brems}= 3\times 10^9\,{\rm sec}\; M_{6}
R_{16}/\dot M_{-4}$. By comparison, the subsonic flow time scale is
$t_{\rm flow}=H_{\rm i}/v_{\rm sh}$, where $H_{\rm i}$ is the scale
height of the shock above the disk.  Thus, because

\begin{equation}\label{tbrems}
{t_{\rm brems}\over t_{\rm w}}=13\;{M_{6}^{3/2}\over\dot
M_{-4}R_{16}^{1/2}}
\left(H_{\rm i}\over R\right)^{-1}
\end{equation}
and $H_{\rm i}\ll R$, the gas cannot radiate appreciably before
reaching the disk.  If we express $R$ in units of the gravitational
radius, this ratio is inversely proportional to the Eddington
luminosity of the black hole, and can become smaller than unity only
for super-Eddington wind accretion or very large radii.  In the
post-shock region above the surface of the disk, the state of the gas
is then almost always dictated solely by the influence of Coulomb
scatterings and conductive heat transport.

Let the temperature, density, and height of this interface region
between shock and disk be $T_{\rm i}$, $n_{\rm i}$, and $H_{\rm i}$,
respectively, where $T_{\rm sh}>T_{\rm i}>T_{\rm d}$ and presumably
$H_{\rm i} \ll R$ (see below). The density is given roughly by the
balance of ram pressure and thermal (surface) pressure:
\begin{equation}
n_{\rm i}=2.8\times 10^9\,{\rm cm}^{-3}\;{M_{6}^{1/2} \dot
M_{-4}\over R_{16}^{5/2}}\left({T_{\rm i}\over 1000\,{\rm K}}\right)^{-1}.
\end{equation}
The  conductive heat flux in the interface region is
$q\sim2.5\times 10^{-6}\,{\rm erg/sec/cm}^2\;(T_{\rm i}/{\rm
K})^{7/2}/(H_{\rm i}/{\rm cm})$ (Frank, King \& Raine 1985, hereafter
FKR, Eq. 3.42). It is not difficult to show that the ratio of the
thermal conductivity time scale $t_{\rm con}=H_{\rm i}\cdot(dE/dV)/q$
to the flow time $t_{\rm i}=H_{\rm i}/v_{\rm sh}$ is
\begin{equation}
{t_{\rm con}\over t_{\rm i}}\approx 0.35\cdot\left(H_{\rm i}\over R\right)
{\dot M_{-4} R_{16}^{3/2}\over M_{6}^{5/2}}\;,
\end{equation}
which in most cases is much smaller than unity if $H_{\rm i}\ll R$,
remains valid even for Eddingtion accretion and large radii, and is
independent of the disk temperature.

An upper limit to the value of the scale height $H_{\rm i}$ is given
by the Coulomb energy exchange length, based on the scattering time
scale $t_{\rm coul}=4.6\times 10^{-25} v_{\rm sh}^5 n_{\rm i}^{-1}
T_{\rm i}^{-1}\, {\rm K\,sec^6\,cm^{-8}}$ (FKR, Eq. 3.32) on which
incoming particles share their energy with the background plasma.
Since this time scale is extremely short, the wind/disk interface
region is extremely thin (even compared to the disk scale height):
\begin{equation}\label{depth}
{H_{\rm i}\over R}\la 2\times 10^{-6}{\dot M_{6}^{5/2}\over \dot M_{-4}
R_{16}^{3/2}}\;.
\end{equation}
Thus, because of the high efficiency of thermal conduction and Coulomb
scatterings, and the relatively large ram pressure, in almost all
cases of interest most of the energy in the flow is deposited directly
in a very thin layer onto the surface of the disk. Since the surface
of the disk (in contrast to its interior) may be only weakly ionized,
the incoming ions might actually penetrate somewhat below the disk
surface until Coulomb interactions start to work and as $H_{\rm i}$ is
so small, most of the energy will be deposited somewhere in or even
beneath the surface layer. Given these conditions, we shall treat the
disk as a large mass and heat sink for the inflowing wind.  In
particular, we shall assume for simplicity that the deposited energy
is radiated as a (local) black body, since the spectral modifications
due to bremsstrahlung emission above this region are expected to be
small (see Eq. \ref{tbrems}).

\section{Modified Disk Equations}
\subsection{Mass and angular momentum conservation}
The basic equations describing the evolution of an accretion disk are
the mass conservation equation and the angular momentum equation (von
Weizs\"acker 1948; Lynden-Bell \& Pringle 1974; see also Equations~5.3
and 5.4 in FKR). In the case of an accretion disk intercepting an
external wind, these have to be modifed to include additional source
terms from the external gas flow.  If the wind produces a mass infall
\dsigw(r,t) per unit area and time with a velocity $\vec u = (u_{\rm
r},u_\phi,u_{\rm z})$ in cylindrical coordinates, we obtain the mass
equation

\begin{equation}\label{masscon}
r {\D\Sigma\over\D t} = -{\D \over \D r} \left(r\Sigma v_{\rm
r}\right) + r \dsigw
\end{equation}
and the angular momentum equation

\begin{equation}\label{angcon}
r {\D\over\D t}\left(\Sigma r^2\omega\right) = -{\D \over \D r} \left(r\Sigma v_{\rm
r} r^2 \omega\right) + {1\over 2 \pi}{\D G\over \D r} + r \dotl\;.
\end{equation}
The other variables used here are the disk surface density $\Sigma$, the disk
radius $r$, time $t$, (local) wind angular momentum flux \dotl$\equiv r\dsigw
u_{\phi}$, disk internal viscous torque $G$=$2\pi r \nu \Sigma r^2
\omega'$, partial radial derivative $'$ = ${\D/\D r}$, internal
disk viscosity $\nu$, and the disk angular velocity
$\omega$=$\sqrt{G_{\rm grav}M/r^3}$. Combining these equations yields 
\begin{equation}\label{dsigw}
{\D \Sigma\over\D t}={3\over r} {\D\over\D r}\left[
r^{1/2}{\D\over\D r}\left(\nu\Sigma r^{1/2}\right) + 
{2\over3}(1-\xi)^{-1}\nuw \Sigma\right] +\dsigw,
\end{equation}
which is the well known partial differential equation for the time
evolution of the surface density of the disk, but now including the
effects of wind infall. The second equation, obtained by combining
Equations~\ref{masscon} \& \ref{angcon}, is the radial velocity of the
accreting gas

\begin{equation}\label{vr}
v_{\rm r}=-{3\over r\Sigma}\left[r^{1/2}{\D\over\D r}\left(\nu \Sigma
r^{1/2}\right)+ {2\over3} (1-\xi)^{-1}\nuw \Sigma\right].
\end{equation}

To permit a straightforward identification of the physical meaning of
the new terms, and to provide a means of comparing the various effects
in a simple manner, we define an external, rotational `viscosity' \nuw\ 
that describes the braking of the disk by the wind:

\begin{equation}\label{nuw}
\nuw = (1-\xi)^2 r^2 {\dsigw\over\Sigma},
\end{equation}
and a parameter $\xi(r,t)$ that describes the angular velocity of the wind
relative to its Keplerian value $v_\phi\equiv r\omega$, such that

\begin{equation}
u_\phi = \xi v_\phi\;.
\end{equation}
For the type of accreted wind discussed in this paper, we have
$-1\le\xi\le1$. The term $(1-\xi)^2$ in Equation~\ref{nuw} ensures that the
viscosity is always positive even if the angular velocity of the
external wind is larger than the Keplerian value. In the case of a
wind flowing out from the center intercepting the disk with $\xi>1$,
the disk would be spun up and the radial disk inflow would be diminished
by the external wind. This is reflected by a negative forefactor
$(1-\xi)$ inside the brackets of Equations \ref{dsigw} \& \ref{vr}. The
only disadvantage of this definition of $\nu_{\rm w}$ is the apparent
singularity at $\xi=1$, which, however, is not real as then also
$\nu_{\rm w}=0$.

In this definition we only consider the viscosity induced by the
azimuthal velocity of the wind. The radial contributions of the
momentum equation are neglected here but are considered in the
energy equation as discussed below.

\subsection{Energy equation}
Any kind of viscosity, which is needed for the angular momentum and
mass transport in the disk, is intimately linked to friction and
dissipative processes. In accretion disk theory, it is usually assumed
that the dissipated energy is radiated locally, though advection
can under certain circumstances carry a significant fraction of the 
flux.  The rate of energy
dissipation is a function of the disk surface density, the viscosity
and the shear rate. For the internal energy dissipation caused by
turbulent viscosity and differential rotation in the disk, the rate of
dissipated energy per unit area and time is (see FKR, Equation~4.25)

\begin{equation}\label{Dd}
D_{\rm d}={1\over2}\,\nu\Sigma(r\omega')^2.
\end{equation}
This is defined as dissipated energy per surface area and as we have 2
sides of the disk we get a factor of 1/2.  Braking of the disk by the
wind will cause a change in angular momentum, lead to friction between
wind and disk and finally to energy dissipation. For now we will
neglect the vertical structure of the disk and wind and treat the
disk-wind interaction zone as an infinitesimally thin sheet in which
the energy is dissipated (see Sec. 2 above). The dissipation rate,
which is the change of energy $\Delta E$ per time $\Delta t$ and
surface area $\Delta A$ due to the friction between the wind and disk,
can then be written as

\begin{eqnarray}
D_{\rm w} &=& {1\over 2}\,{\Delta E\over \Delta t\;\Delta A} \nonumber\\
  &=& {1\over 2}\,{2 \pi r (\Delta G_{\rm w}/r)\over \Delta t\;\Delta A}\nonumber\\
  &=& {1\over 2}\,(v_{\phi}-u_{\phi})^2\dsigw\;,
\end{eqnarray}
using the torque 
\begin{equation}
\Delta G_{\rm w} = r \dsigw (v_\phi-u_\phi)\Delta A
\end{equation}
exerted by the infalling wind on the accretion disk on a time scale 
\begin{equation}
\Delta t = 2 \pi r/(v_\phi-u_\phi)\;,
\end{equation}
given by the azimuthal velocity of the wind relative to the disk.
Thus we obtain the final form for the dissipation rate and can express
this in terms of the rotational viscosity $\nu_{\rm w}$ defined above:

\begin{eqnarray}\label{Dw}
D_{\rm w}&=& (1/2)\,r^2 \omega^2 (1-\xi)^2\, (\dsigw/\Sigma)
\,\Sigma\nonumber\\
&=&(1/2)\, \nuw\, \Sigma\, \omega^2\;.
\end{eqnarray}
Comparison of Equations~\ref{Dw} and \ref{Dd} shows that our definition
of \nuw{} (Equation~\ref{nuw}) is indeed justified.

In addition to the dissipative heating by rotational and turbulent
friction we have to take the ram pressure of the wind into account,
which results in a heating rate of a form similar to Equation~\ref{Dw}. We
shall consider both the $u_{\rm r}$ and $u_{\rm z}$ components of 
the wind ramming into the
disk. It is assumed that the radial velocity of the disk is much
smaller than the wind velocity $v_{\rm r}\ll\sqrt{u_{\rm r}^2 + u_{\rm
z}^2}$ and that the radial pressure gradient induced by the wind is
negligible for the disk structure. If all the kinetic energy associated
with $u_r$ and $u_z$ is converted into heat and thus contributes to the 
total dissipation rate, we have an additional term

\begin{equation}\label{Dram}
D_{\rm ram}={1\over2}\dsigw \left(v_{\rm r}^2+v_{\rm z}^2\right).
\end{equation}
The total dissipation rate per surface area is then given by 
\begin{equation}\label{dissip}
D=D_{\rm d}+D_{\rm w}+D_{\rm ram}.
\end{equation}
One should note that in any realistic case, the wind properties will be
asymmetric with respect to the two sides of the disk: one would have
to split Equations~\ref{Dw} and \ref{Dram} accordingly and introduce
separate parameters $\xi_{1|2}$, $\dsigw_{1|2}$ and $\nuw_{1|2}$ and 
obtain asymmetric dissipation rates. We do not want to introduce this
complication into our model at the current stage.

\subsection{The vertical structure}
To solve the disk equations and obtain the evolution of $\Sigma$ we need
to know the radial dependence of the turbulent viscosity $\nu$. This is
usually done by introducing the Shakura \& Sunyaev (1973) $\alpha$ parameter,
defined such that 

\begin{equation}
\nu=\alpha c_{\rm s} z_0\;,
\end{equation}
where $c_{\rm s}$ is the local sound speed, $z_0$ is the scale height of the disk 
and $\alpha$ is the dimensionless viscosity parameter, 
usually between 0 and 1. This has the benefit that
the viscosity can be expressed in terms of local properties of the disk but has
the disadvantage that one needs to solve its vertical structure.

The disk scale height can be obtained by demanding pressure equilibrium between 
internal gas pressure, gravitational attraction in the z-direction and 
external ram pressure from the wind. Substituting the gradient $\D/\D z$ 
with the scale height $z_0$, we obtain for a gas-pressure dominated disk
with sound speed $c_{\rm s}$

\begin{equation}
z_0 = {c_{\rm s}\over\omega}\left(\sqrt{\left({1\over2}{\dsigw\over
\Sigma\omega}{u_{\rm z}\over c_{\rm s}}\right)^2+1}-{1\over2}{\dsigw\over
\Sigma\omega}{u_{\rm z}\over c_{\rm s}}\right).
\end{equation}
This means that the disk is substantially squeezed if $u_{\rm
z}\dsigw/\Sigma\gg 2 c_{\rm s} \omega$; if $u_{\rm z}$ is of the order
of the Keplerian velocity (i.e., $u_{\rm z}\sim r\omega$) then we have the
condition 

\begin{equation}
2\Sigma/\dsigw \ll (c_{\rm s}r)^{-1}
\end{equation}
for the time scale involved in order to compress the disk.  The left
hand side is the time scale of the wind interception and the right
hand side the {\em global} thermal time scale of the disk.

To complete our set of basic equations, we need to specify the central
temperature in the disk to get the sound speed.  We assume that the
energy dissipated by viscous processes {\em in the disk} diffuses
outwards from the central plane by radiative energy transport in an 
optically thick medium  (with optical depth $\tau\gg1$)
in local thermal equilibrium.  This will give us an energy flux of
dissipated energy in the disk at its surface of

\begin{equation}\label{Fd}
F_{\rm d}(z_{0})=D_{\rm d}=-{4\sigma\over 3}{\Delta T(z_0)^4\over \Delta\tau},
\end{equation}
the $\Delta$ here signifying changes in the corresponding quantity between
the disk's midplane and its surface.
The total energy flux will then be the sum of the energy dissipated in
the disk and the energy dissipated by the impact of the wind on the
surface of the disk. If this is radiated immediately by black body
radiation with an effective temperature $T_{\rm eff}$, we have

\begin{equation}\label{teff}
F_{\rm tot}(z_0)=\sigma T_{\rm eff}(r,t)^4 = D(r,t),
\end{equation}
and we can set $T(z_0)=T_{\rm eff}$. With $\Delta T \equiv T(z_0)-T(0)$ and 
$\Delta \tau \sim \tau$, we can solve Equation~\ref{Fd} for $T(0)$, getting

\begin{equation}
T(0)^4={1\over\sigma}\left[\left({3\tau\over4}+1\right)D_{\rm
d}+D_{\rm w}+D_{\rm ram}\right].
\end{equation}
This equation reduces to the usual equation for the central temperature if 
$D_{\rm w}\sim0$, $D_{\rm ram}\sim0$ and $\tau\gg1$, but it also allows a
situation where the dissipation is dominated completely by the impact of
the wind. This will increase the central temperature and make the
temperature gradient shallower and in the extreme case could lead to a
quasi-isothermal disk; in the cases discussed in this paper, this
does not happen. 

With the central temperature at hand we can now evaluate the pressure
\begin{equation}
P={\rho k_{\rm b} T\over \mu m_{\rm p}}
\end{equation}
and the sound speed
\begin{equation}
c_{\rm s}=\sqrt{P/\rho},
\end{equation}
if we solve the second order partial differential equation for
$\partial\Sigma/\partial t$ 
(Equation~\ref{dsigw}).  We only need to specify the opacity. Assuming
that we are in the outer and cooler region of the disk, we may 
consider simply the contribution from free-free absorption which gives us

\begin{equation}
\kappa = 6.6\cdot10^{22} {\rm cm}^2/{\rm g} \left({\rho\over {\rm 1\;g/cm}^3}\right) 
\left({T\over{\rm 1\;K}}\right)^{-7/2}.
\end{equation}
This opacity will become progressively less accurate as we move to very cold 
or very hot disks, but we ignore this trend here, since we are primarily 
interested in a basic understanding of the disk evolution and the main effects. We
hope to include a more elaborate opacity function for more realistic
calculations in a future version.


The complete set of equations can now be solved numerically. For each time
step we calculate the radial gradients in $\Sigma$ on the right-hand
side of Equation \ref{dsigw} to get $\partial \Sigma/\partial t$ and an
estimate for $\Sigma$ at the next time step using a simple Runge-Kutta
algorithm with adaptive step size. The vertical structure has to be
calculated at every step and because the height is no longer a simple
function of $c_{\rm s}$ and $\omega$, we have to solve this set of
equations implicitly. Fortunately, $c_{\rm s}$ changes only slowly so
one can use the sound speed from the previous step as a first guess and 
achieve convergence very quickly.

\subsection{Spectrum}
To get a rough spectrum from the accretion disk, we have integrated
black body spectra with the local effective temperature $T_{\rm
eff}(r)$ over the entire disk surface. Following our discussion in
Sec. 2, we consider $T_{\rm eff}$ to represent the total dissipation
rate, as indicated in Equations \ref{dissip} \& \ref{teff}. Here, we
have neglected all vertical transport processes that might modify the
disk spectrum.  Mainly because of the difficulties in comparing disk
spectra to observations and the uncertainties involved in the vertical
disk structure itself, the theory for radiation transport in disk
atmospheres is not well developed, especially in the case of AGN
disks. While Sun \& Malkan (1989) find that the pure black body
integration can explain AGN spectra sufficiently well, Laor \& Netzer
(1989) have used a simplified treatment to account for the effects of
the vertical structure. The changes to the calculated spectrum are,
however, so small, especially compared to the scarce data for the
Galactic Center, that inclusion of these effects would neither be
opportune nor testable.  For simplicity we also calculate our spectra
for only one inclination angle $i=60^\circ$ (i.e. $\cos i=0.5$). The
spectra we show should therefore be considered only as a first order
approximation and one should bear in mind that orientation effects,
relativistic effects, limb-darkening, and propagation effects will
lead to a somewhat different appearace.

\section{Numerical Solutions}
As an example of the evolution of a disk with wind infall, we choose
parameters that might roughly be appropriate for the Galactic Center,
i.e.  a black hole mass of $M_\bullet=10^6M_{\sun}$, an accretion rate
for the fossil disk of $\dot M_0 = 10^{-8}M_{\sun}$/yr, a wind infall
rate of $\dot M_{\rm w}=10^{-4}M_{\sun}/$yr, and an $\alpha$ of
$10^{-4}$ (e.g. Falcke \& Heinrich 1994).  For simplicity in this
first series of tests, we will correlate the vertical wind velocity
directly to its radial velocity, such that

\begin{equation}
u_{\rm z}=\sqrt{1-\xi^2}v_{\phi}\;\;{\rm  and}\;\; u_{\rm r}=0.
\end{equation}

The boundary conditions for the innermost ring of the disk ($r_{\rm
in}=6R_{\rm g}\equiv 6G_{\rm grav}M_\bullet/c^2$) are such that we
impose $\partial \Sigma/\partial t = 0 $ and $\Sigma=0$ to account for
the presence of the black hole. The outer boundary (at $r=r_{\rm out}$)
can be free (using only the one-sided derivative) or be fixed to a
certain $\dot M(r_{\rm out})$ by virtue of Equation~\ref{vr} and $\dot
M(r)=-2\pi r v_{\rm r}\Sigma$, thus setting the square bracket in
Equation~\ref{dsigw} to $\dot M(r_{\rm out})/6\pi$ at $r=r_{\rm out}$. The
radial velocity is also limited to not exceed the free-fall
velocity. Even though we are here discussing the situation pertaining
to a black hole, we ignore any relativistic effects, which are not 
critical as long as the wind affects mainly the outer portions 
of the disk. During the calculations we check that the total mass in 
the system is indeed conserved.

\subsection{Validation of the code}
To validate our code and test its reliability, we first calculate the
disk evolution in two scenarios without wind infall to compare the surface
density profiles with those of the stationary solution.

\paragraph{Empty Disk}\label{empty}
We start with an almost empty, very low $\Sigma$ disk and feed it with
a very high accretion rate close to the Eddington value of $\dot
M=10^{-2}M_{\sun}$/yr at $r_{\rm out}$ until an equilibrium state is
reached. Figure \ref{fig-empty} shows how the accreted matter quickly
fills the almost empty disk and then slowly approaches the analytical
Shakura \& Sunyaev solution for the steady state (upper panel in
Fig. \ref{fig-empty}). Also, $\dot M$ approaches a constant value for
all $r$.

\placefigure{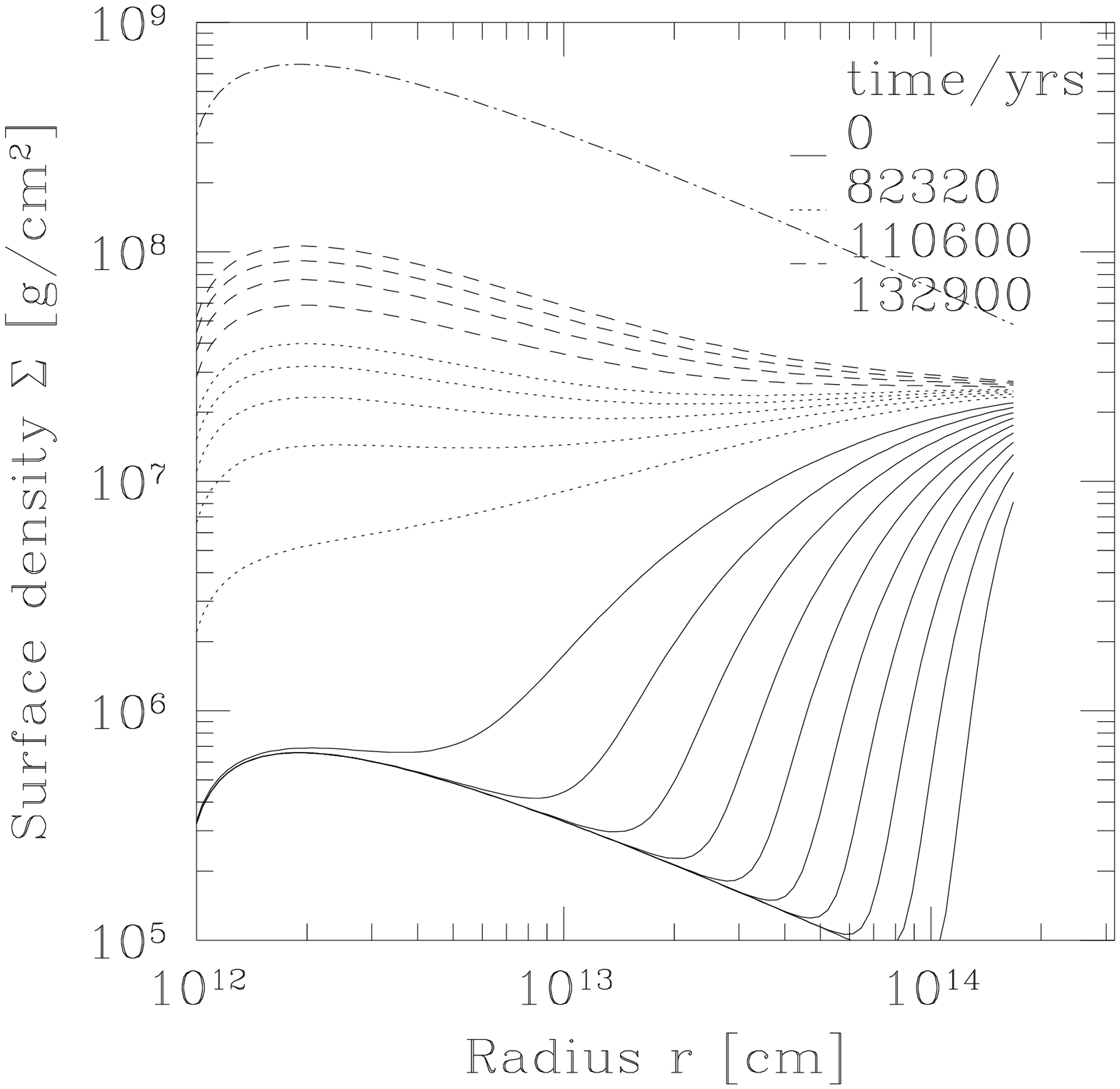}

\placefigure{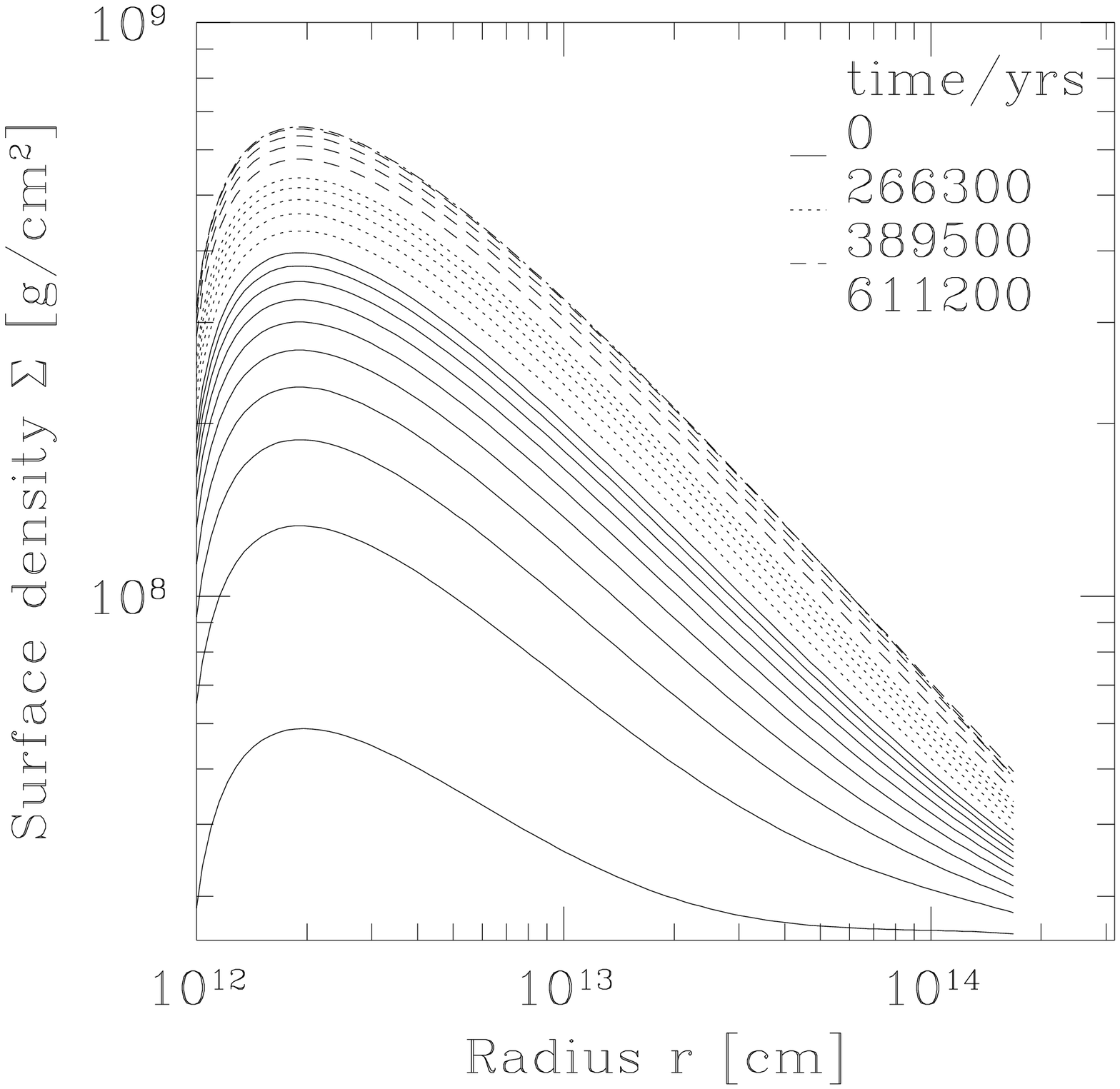}

\paragraph{Gaussian Ring}\label{gauss}
In a second test, a textbook example, we start with a ring with a
Gaussian surface density distribution in $r$ and no accretion from the
outer boundary. Figure \ref{fig-gauss} shows that the ring evolves
towards the stationary solution with constant $\dot M$. In the outer
regions, matter is actually flowing out, carrying away the angular
momentum; only there one can see a deviation from the stationary
solution. In the later evolution (not shown here), the disk runs out
of fuel with a decreasing surface density and size but it retains its
basic shape.

\placefigure{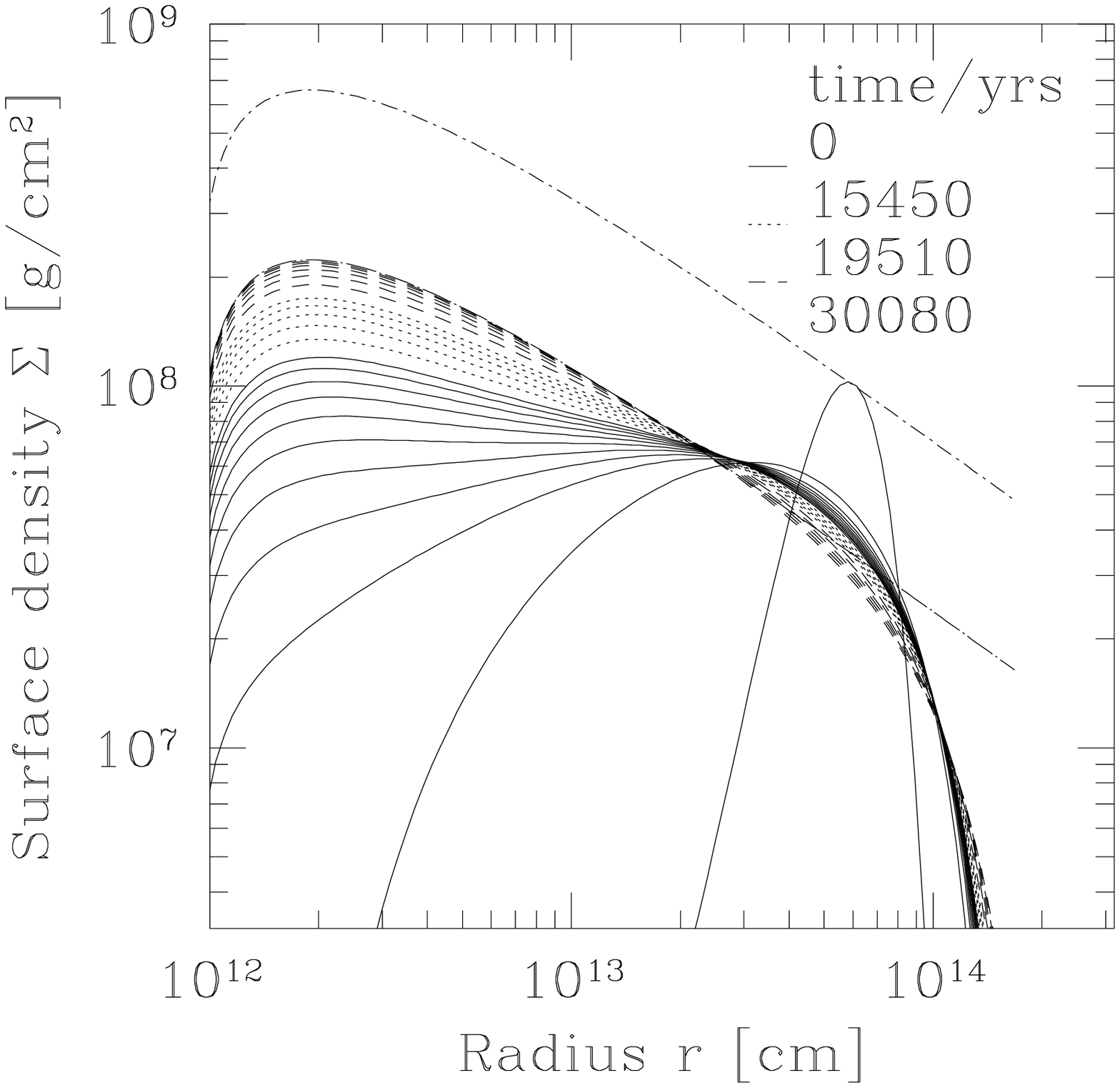}

\placefigure{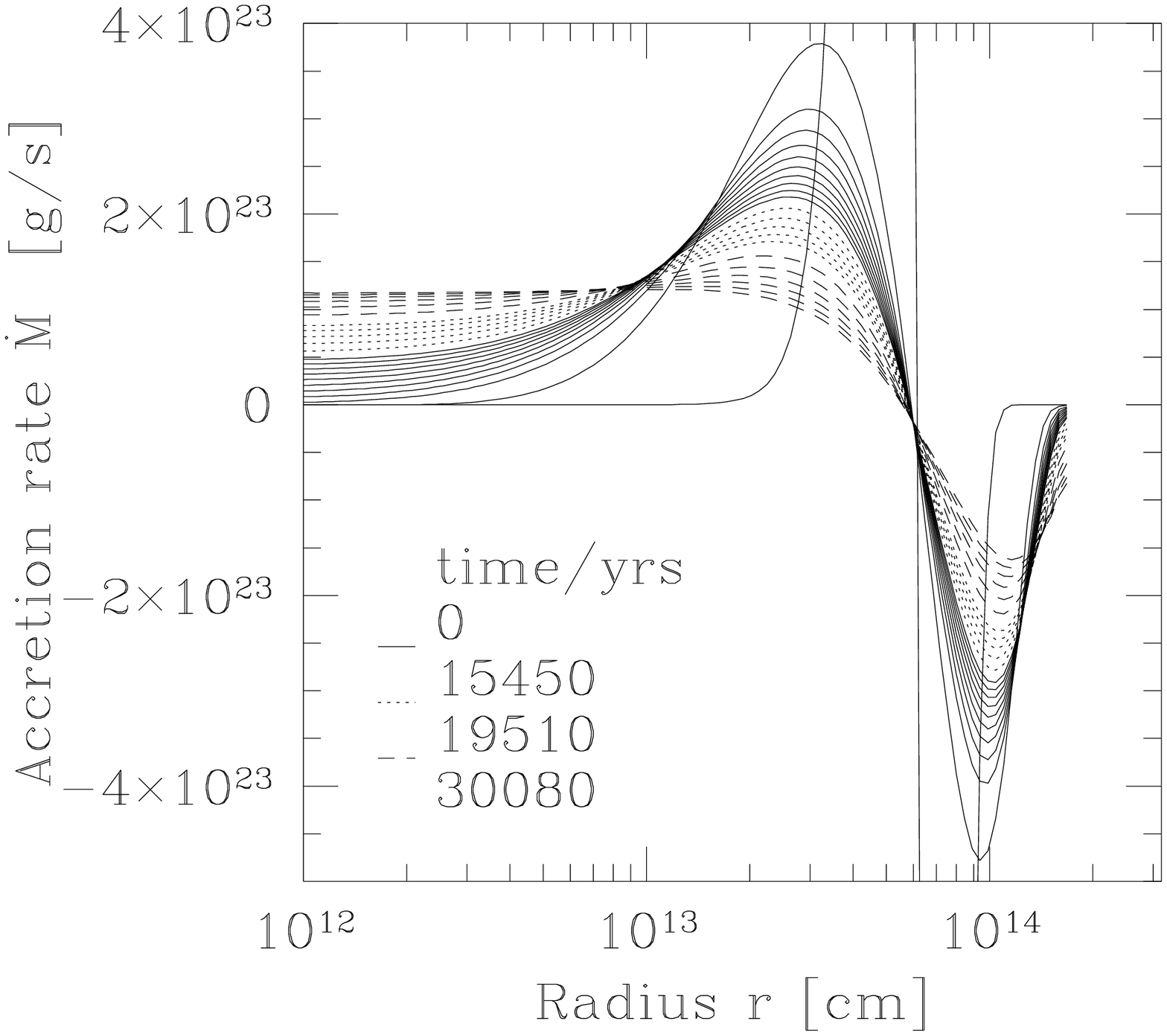}

The agreement between the analytical and numerical solutions demonstrates
that our code produces reliable results.

\subsection{Disk evolution with a large scale wind}
In the first scenario where we allow a wind to fall onto the
disk we assume that the disk intercepts the inflow over a fairly 
large region, extending out to a radius $r=90000 R_{\rm g}
\simeq10^{16}$cm. In the absence of
detailed hydrodynamical simulations, we do not yet know what 
the radial distribution of the wind is and we here consider the case of a
constant mass infall per unit area, which means that the impact of the
wind is felt most strongly at the largest radii, but is 
still present in the inner regions.  More realistic distributions 
will be discussed in the second paper, where we will use the 
results of numerical 3D wind calculations. 

\paragraph{Maximum Angular Momentum}
Figure \ref{fig-wide.xi=1} shows the evolution of a fossil disk
intercepting a wind with constant \dsigw\ over a large collecting area,
as described in the previous paragraph.
Hence, the largest fraction of the wind is intercepted at large radii,
which leads to a significant increase of the surface density there.
Because the viscous time scale (starting with the assumed
$\alpha=10^{-4}$) is too long for a radial
transport of the matter --- the viscosity increases only by a factor of
three during the simulation --- the modest increase in $\Sigma$ at smaller
radii is due mostly to the local wind infall. After many
time steps, the pile-up in the outer disk starts to become less 
pronounced due to radial accretion and the entire disk begins to 
converge towards the stationary solution. The
time scale for an increase in surface density and luminosity is larger
than about $10^6$ yrs. We note that the dynamical time scale 
($\sim 20$ years) for the disk calculated from its rotation velocity 
is $10^5$ times shorter than the time scale for the disk structure 
to change in response to the wind.  Hence, the most notable change 
is expected to be the spectrum. The wind infall leads to a local 
increase in the disk accretion rate and the production of an 
infrared (IR) bump --- the spectrum becomes almost flat.

\placefigure{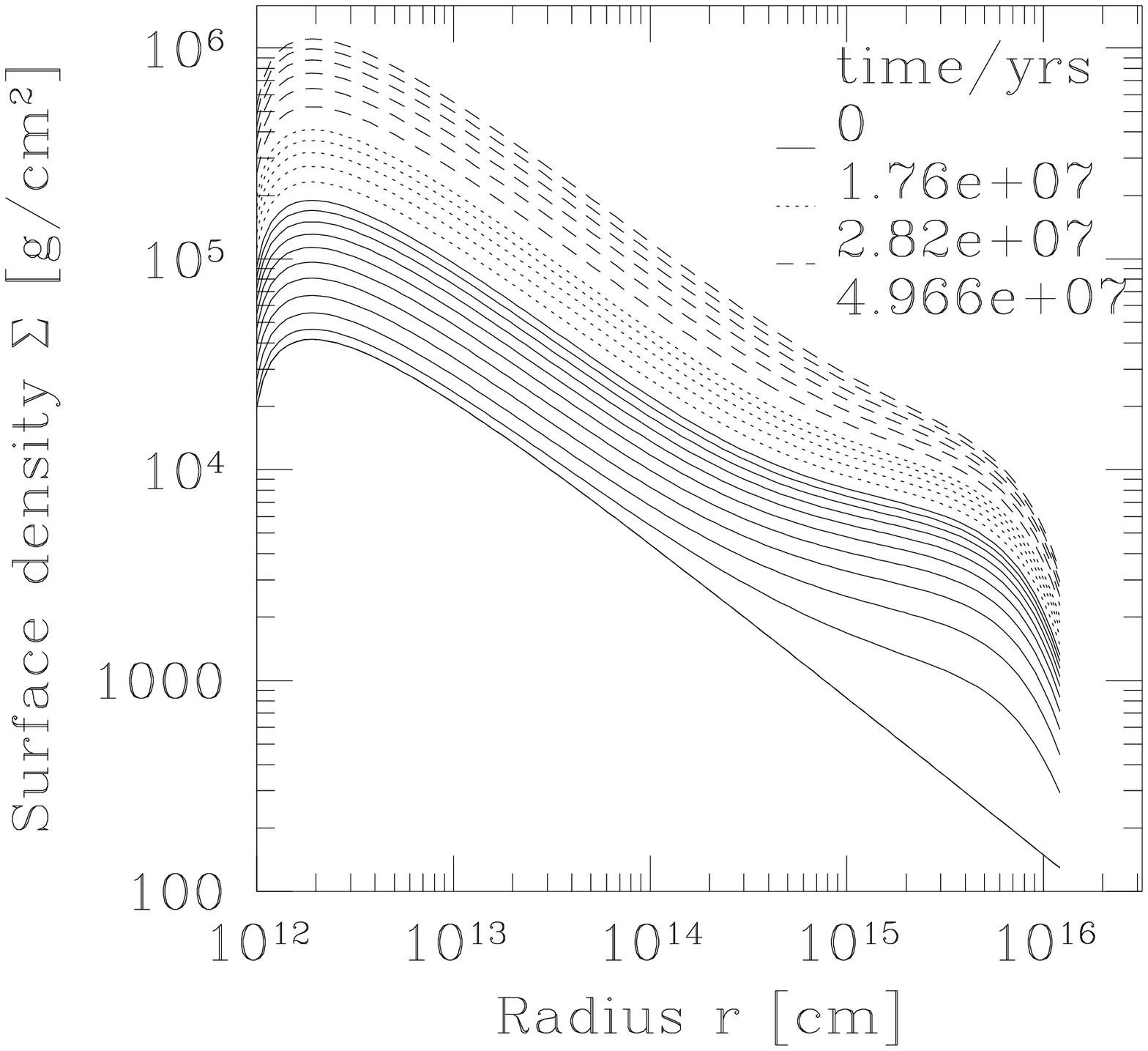}

\placefigure{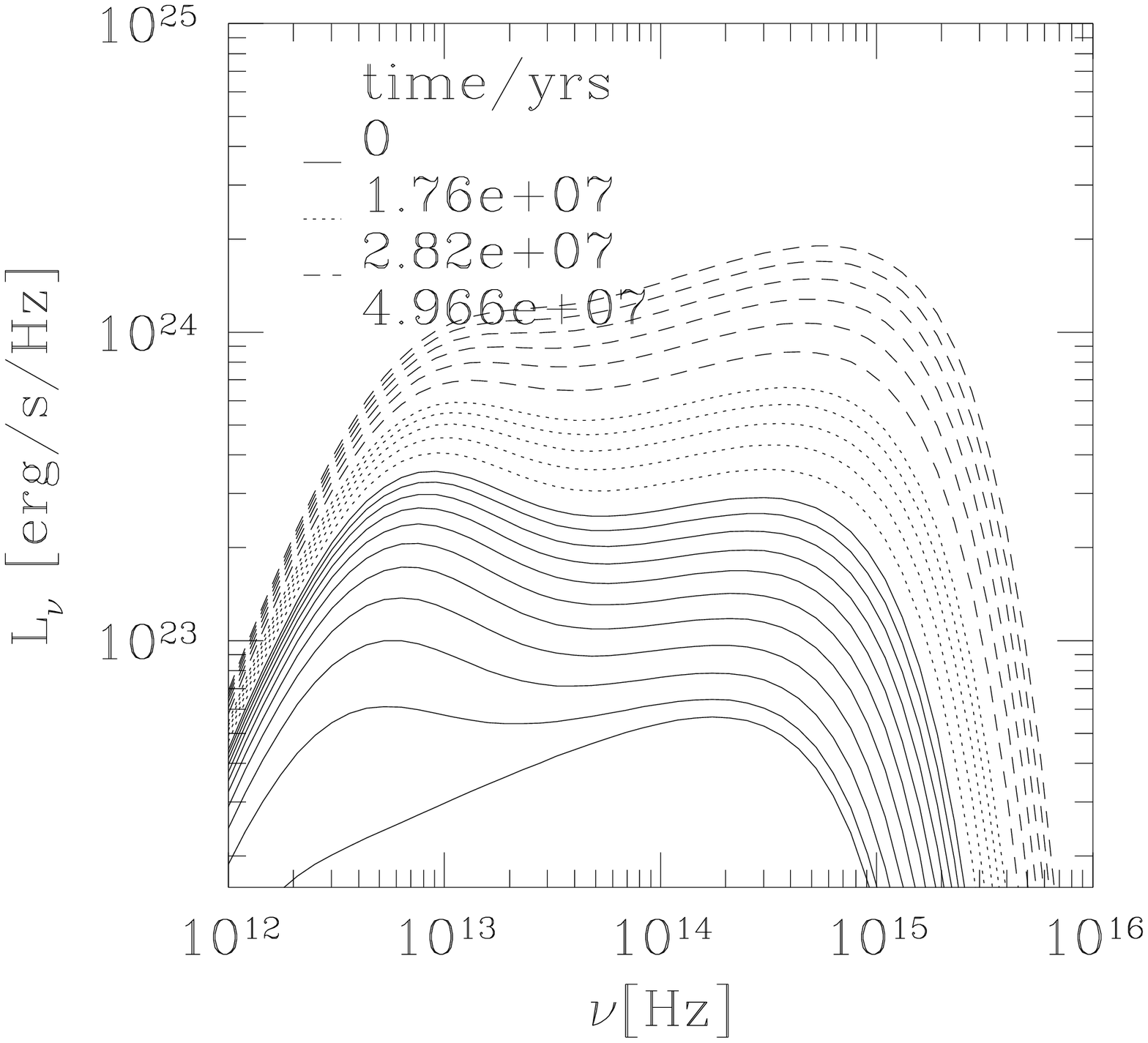}

\paragraph{Hurricane} If we simulate a ``hurricane'' (Fig. 
\ref{fig-hurricane}), where all the mass is deposited within an 
annulus (bound by the radii $30000-90000\;R_{\rm g}$), the
mass piles up in the outer region with a gradual transfer toward 
smaller radii over a $10^8$-yr time scale.  Again, this reflects
the large viscous time scale in the disk assuming that $\alpha=
10^{-4}$ remains constant over this period, and is similar to 
the `empty disk' scenario discussed above.

\placefigure{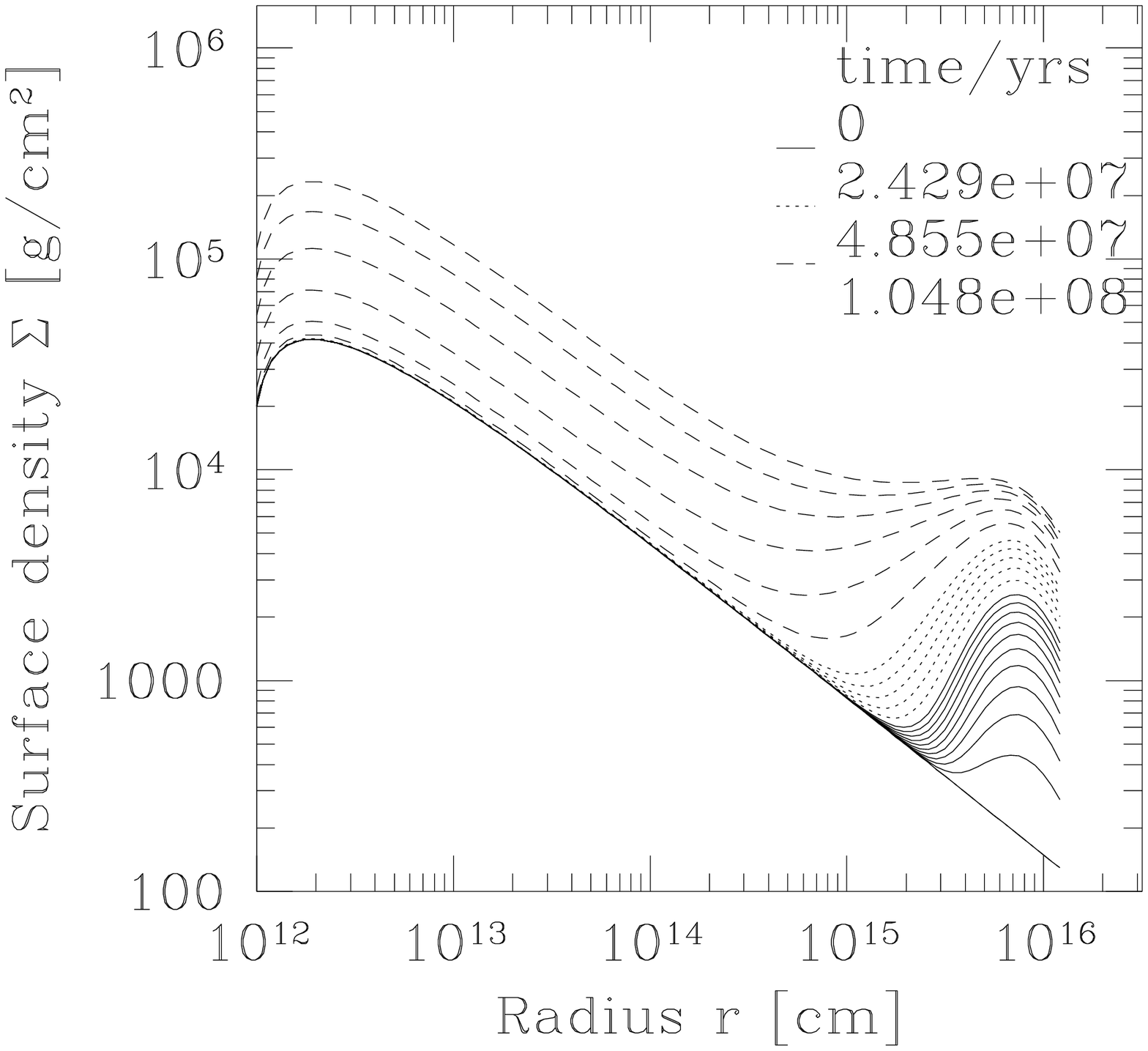}

\placefigure{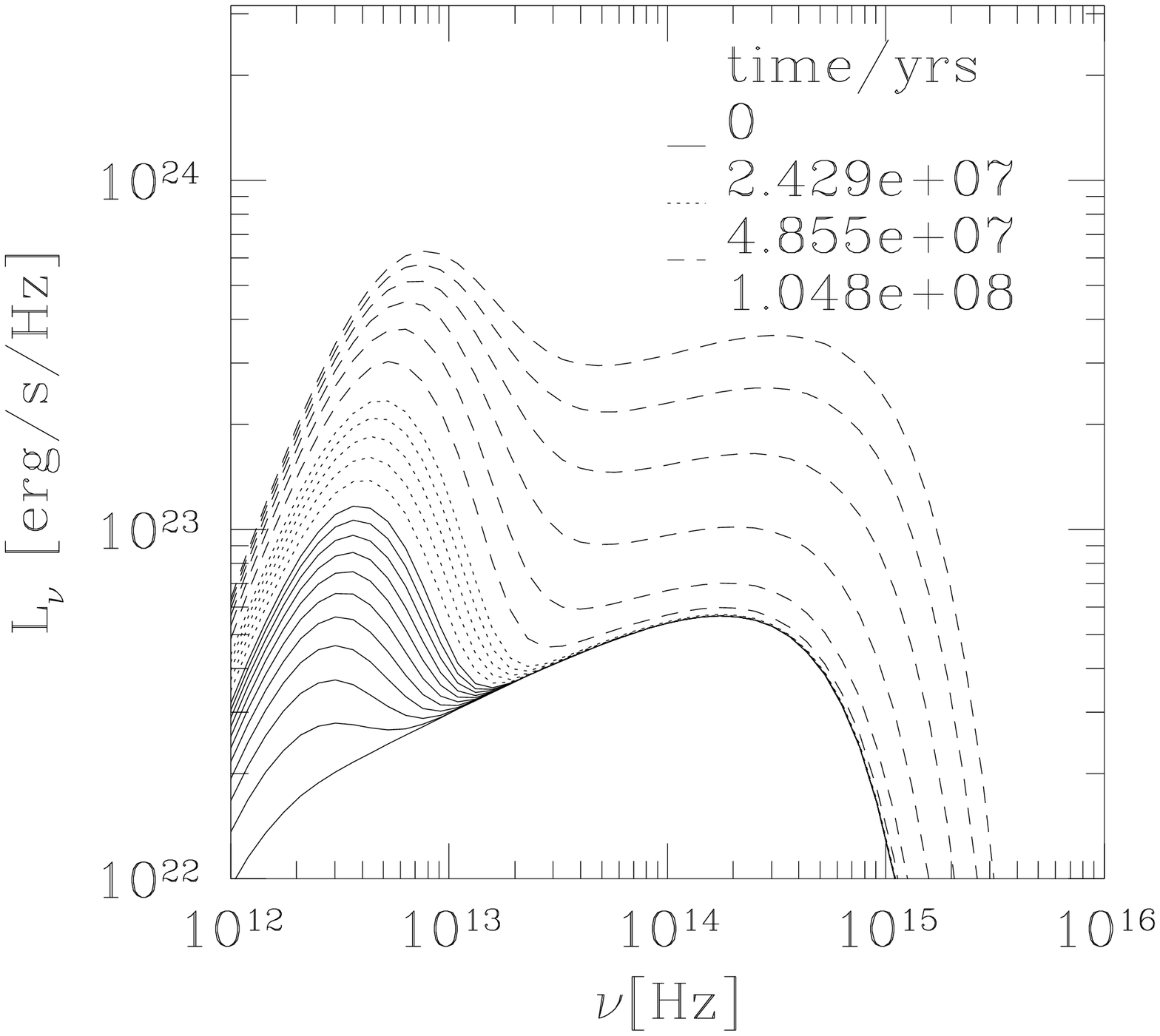}


\paragraph{Zero Angular Momentum}
The time scale for the transfer of mass inwards toward smaller radii
shortens dramatically when the infalling wind carries sub-Keplerian
angular momentum.  The importance of the variations in angular
momentum and velocity distributions in the wind is highlighted in
Figure~\ref{fig-wide.xi=0}, where we have repeated the previous
calculation (shown in Fig. 3) for a free-falling wind with zero
angular momentum ($\xi=0$). The IR bump jumps almost immediately to
the value that corresponds to the high accretion rate in the wind. The
reason for this is that the non-zero vertical impact velocity of the
wind and the strong friction between wind and disk lead to a
rapid dissipation of energy and rapid radial accretion. The deposition
of mass with zero angular momentum in the disk also leads to a
compression of the disk in the radial direction (the outer rings move
inwards as they become sub-Keplerian) and results in an increase in
surface density beyond the value induced by the increased mass
deposition through the wind alone. Also the viscosity increases much
more strongly.

\placefigure{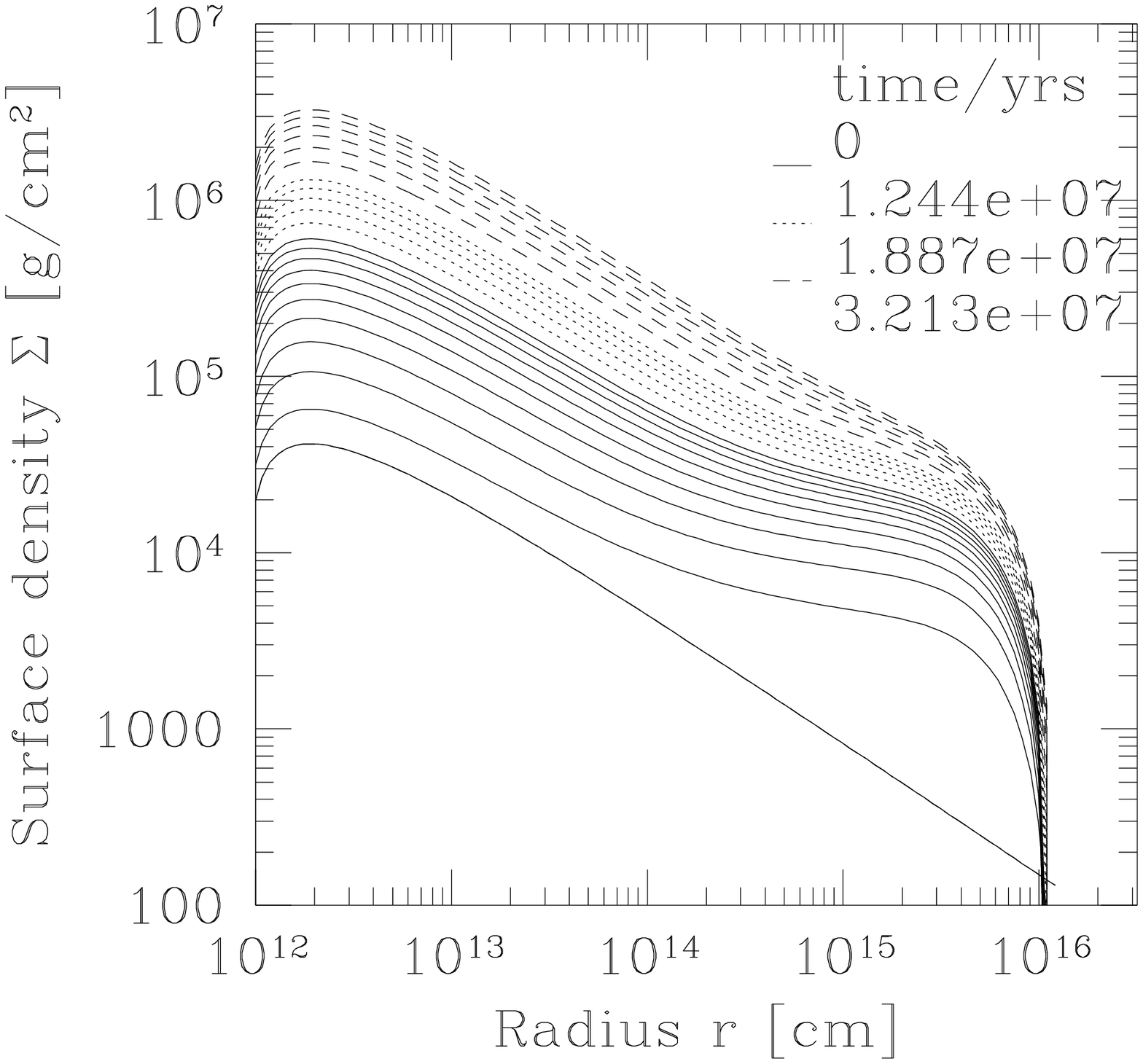}

\placefigure{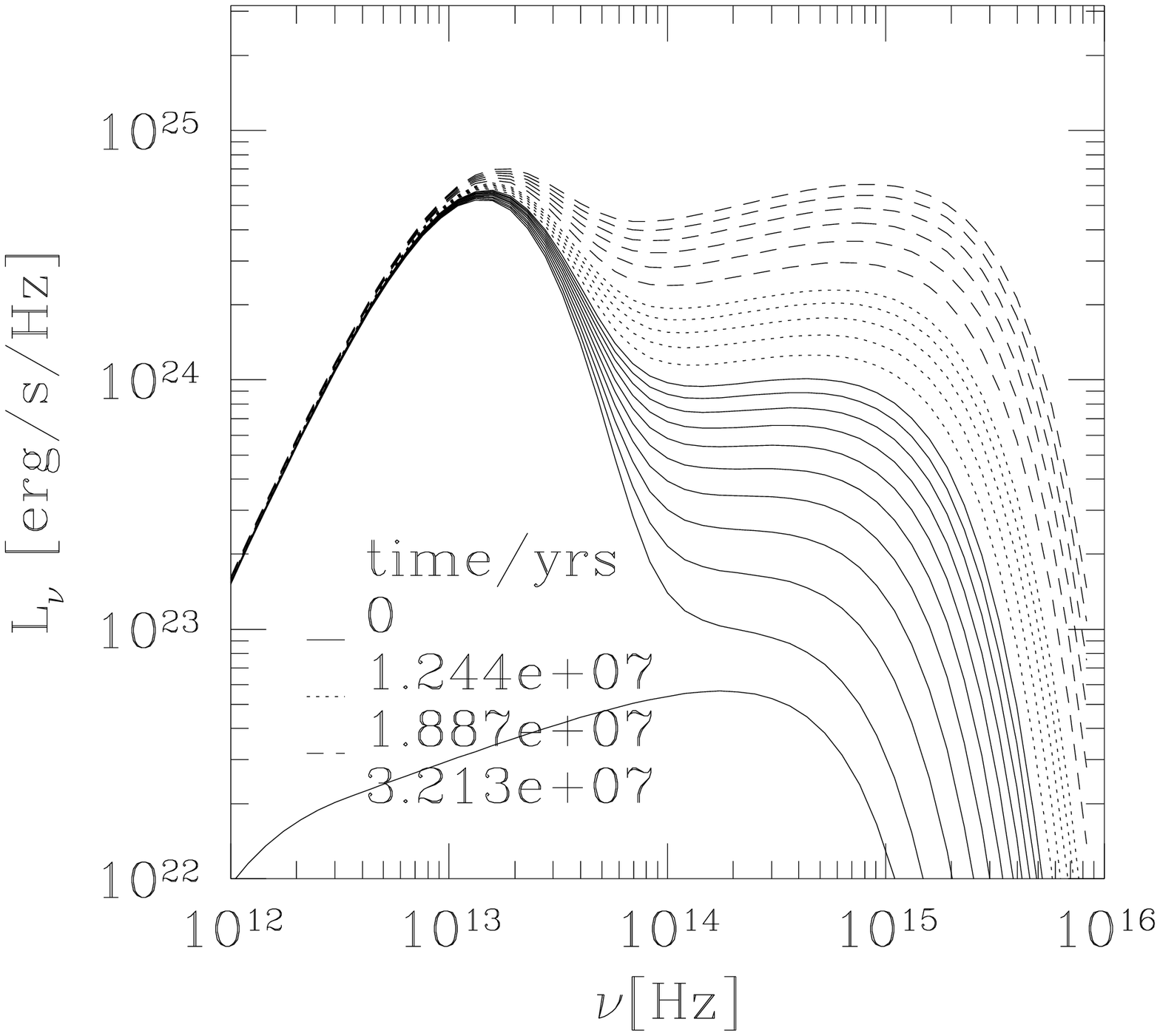}

\subsection{Disk evolution with a small scale wind}
In some systems, the wind flowing into the inner disk region may carry
very little specific angular momentum.  In this case, because the circularization
radius can be significantly smaller than the fossil disk's outer radius, the impact
of the infalling gas on some portions of the disk can be much larger because
the wind is now restricted to a smaller disk surface area and hence accounts for
a significantly larger local flow rate.   We have therefore repeated the 
above calculations, but now with a reduced interception region designated by
a radius 100 times smaller than used in the previous section.  This leads to a much
stronger effect on the disk because of the higher \dsigw\ 
and the correspondingly shorter time scales.

\paragraph{Maximum Angular Momentum}
For a wind with the local Keplerian velocity (Fig. \ref{fig-narrow.xi=1}),
the disk reacts almost immediately at all radii because the
wind infall dominates the radial accretion rate everywhere. Still, the
infall is more prominent in the outer region in the beginning,
yielding an almost radially constant surface density. As soon as the viscous
mass transport has caught up, the surface density profile approaches
the {\it stationary solution}. The mass transport at the outer wind
radius is bi-directional, like in the gaussian density distribution
(Sec.~\ref{gauss}), so that beyond the wind accretion radius matter
is flowing outwards and not inwards. The spectrum is immediately
dominated by emission from the outer region because the kinetic energy
of the wind (in $u_{\rm z}$) is dissipated immediately, but since the
interaction region of the disk is so small, we do not see two bumps. After viscous
processes have transported the higher accretion rate inwards, the
spectrum simply broadens somewhat towards higher frequencies.

\placefigure{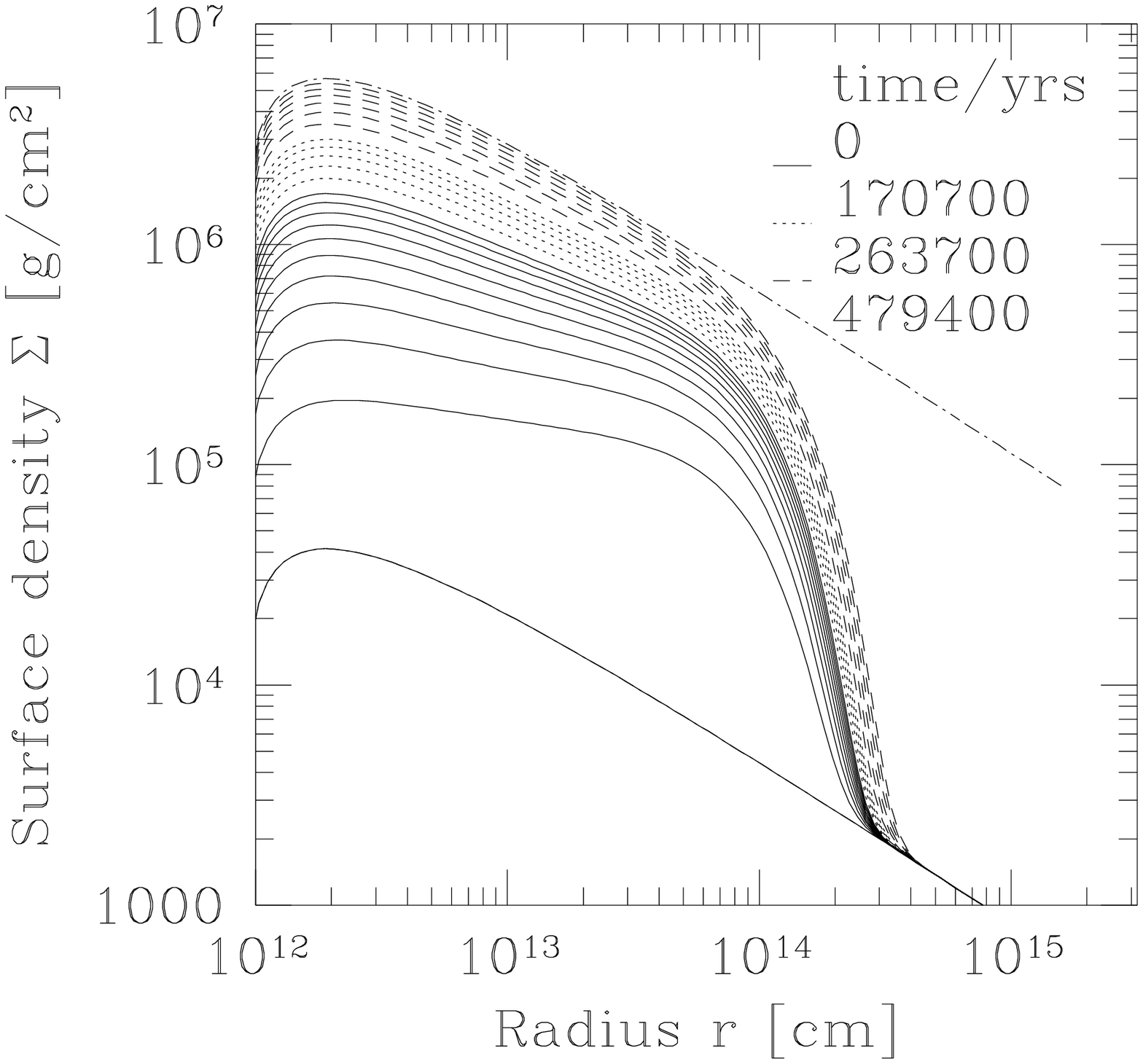}

\placefigure{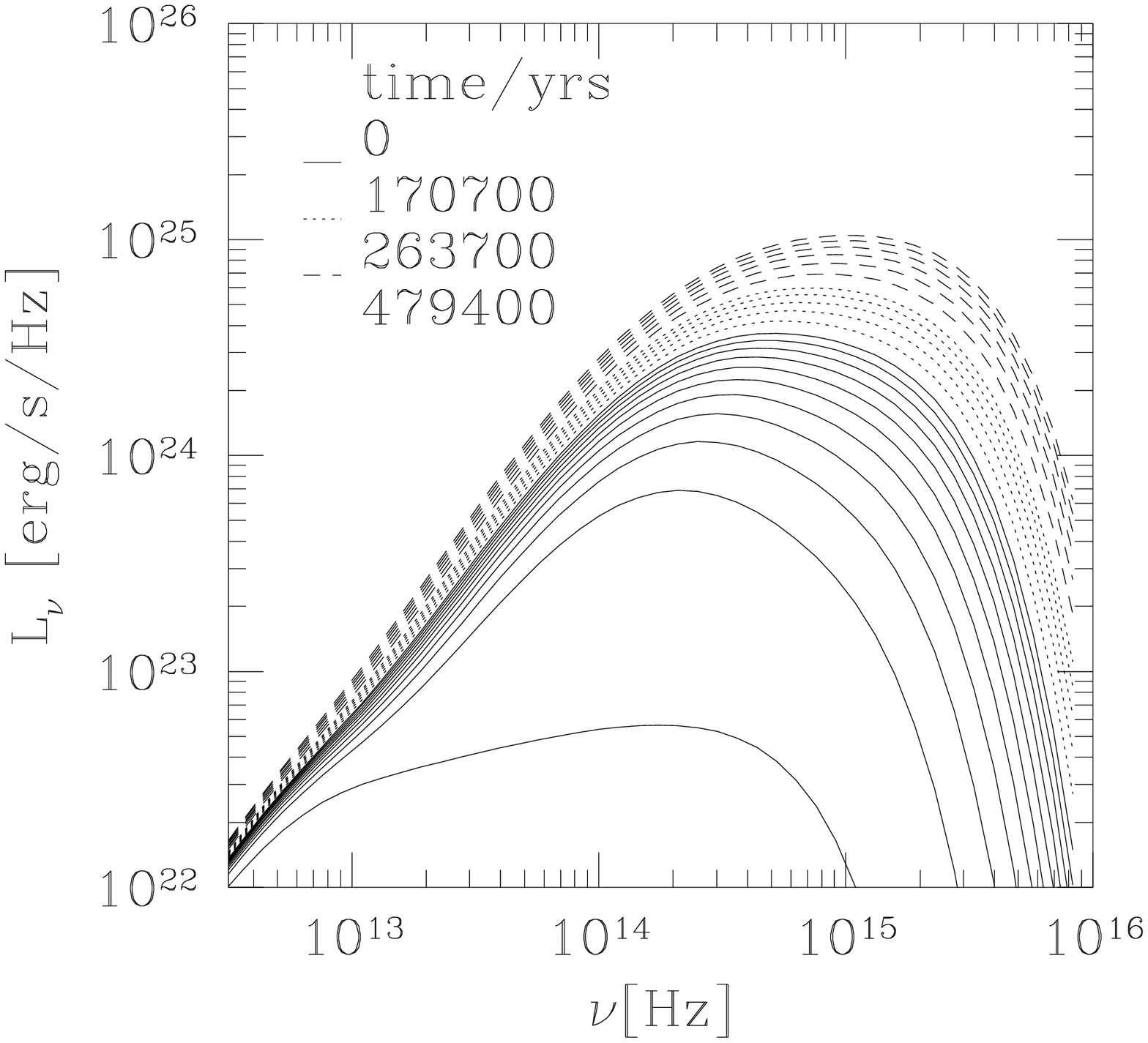}

\placefigure{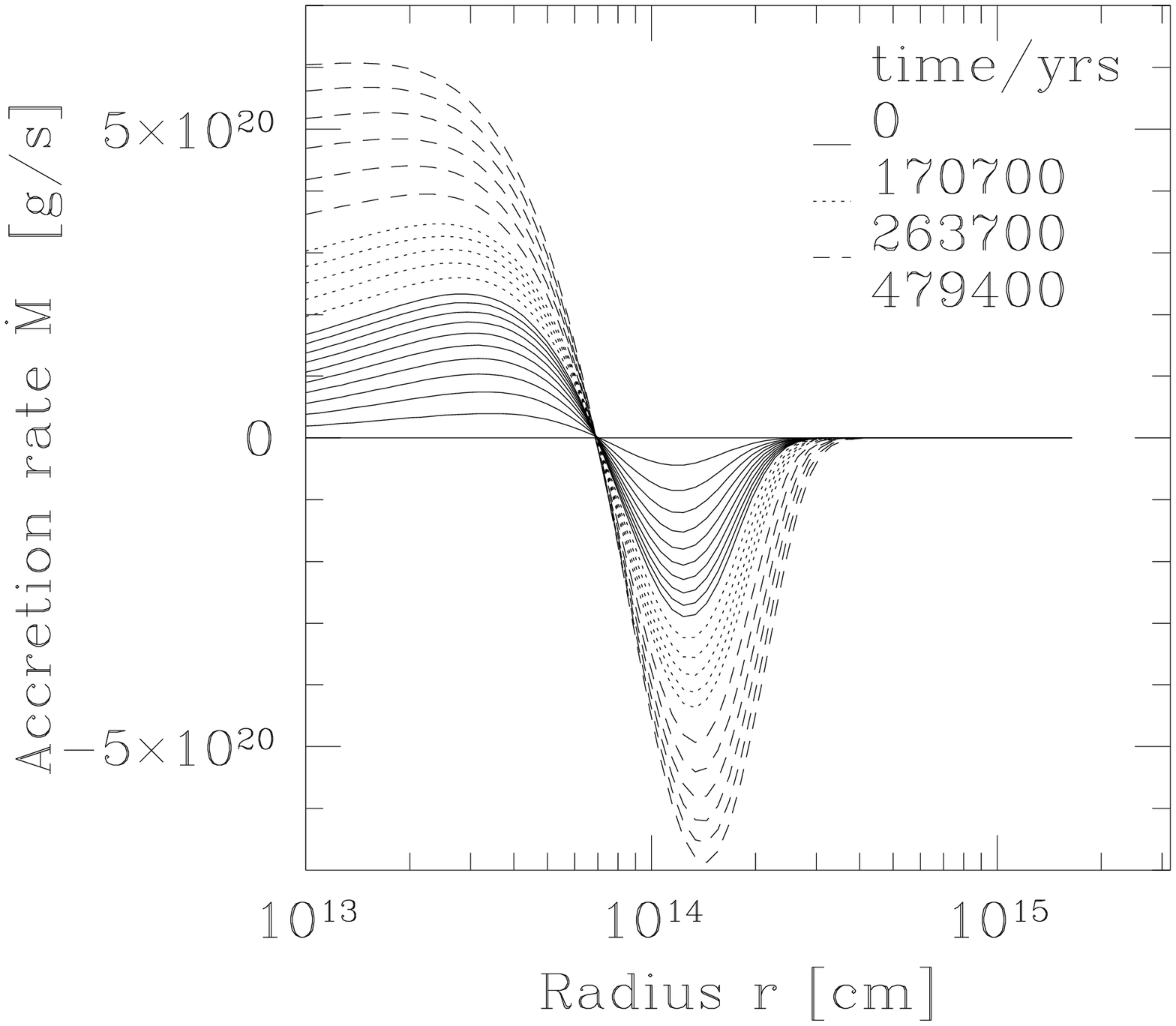}

\paragraph{Zero Angular Momentum}
The situation becomes even more extreme if the wind has no angular
momentum. Within $\sim1000$ years (which is still much longer than the
dynamical time scale) the incorporated mass in the outer region becomes
equal to the stored mass in the disk. Because of the low angular
momentum, the mass falls into a lower orbit within this short
period of time, leaving a gap behind until the fossilized angular momentum
in the disk stops a further free fall. The accretion flow is now
purely inwards and after it is stabilized continues to grow on a
longer time scale ($10^5$ yrs). The spectrum is immediately dominated
by a single Planck-spectrum from the outer region.

\placefigure{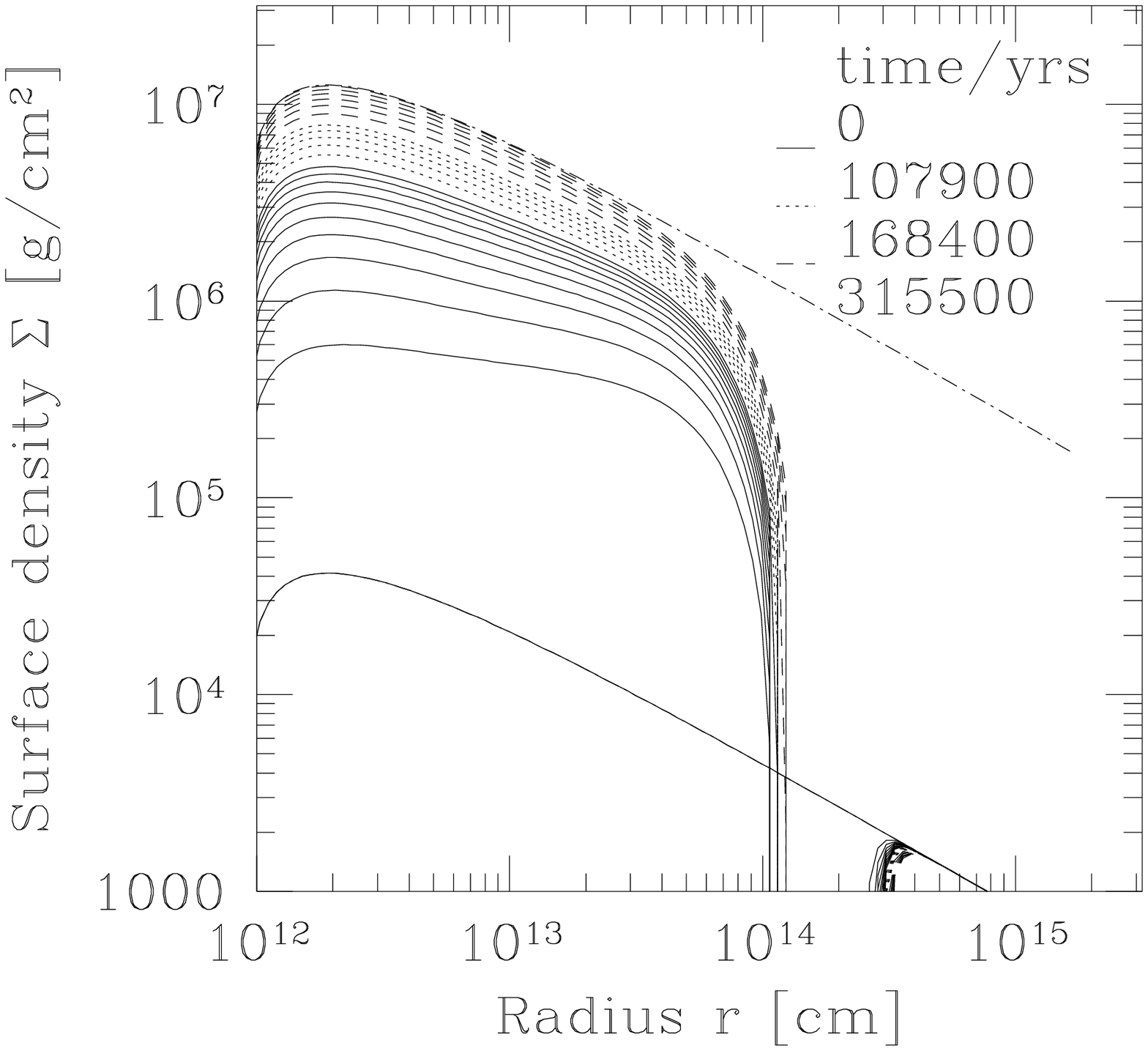}

\placefigure{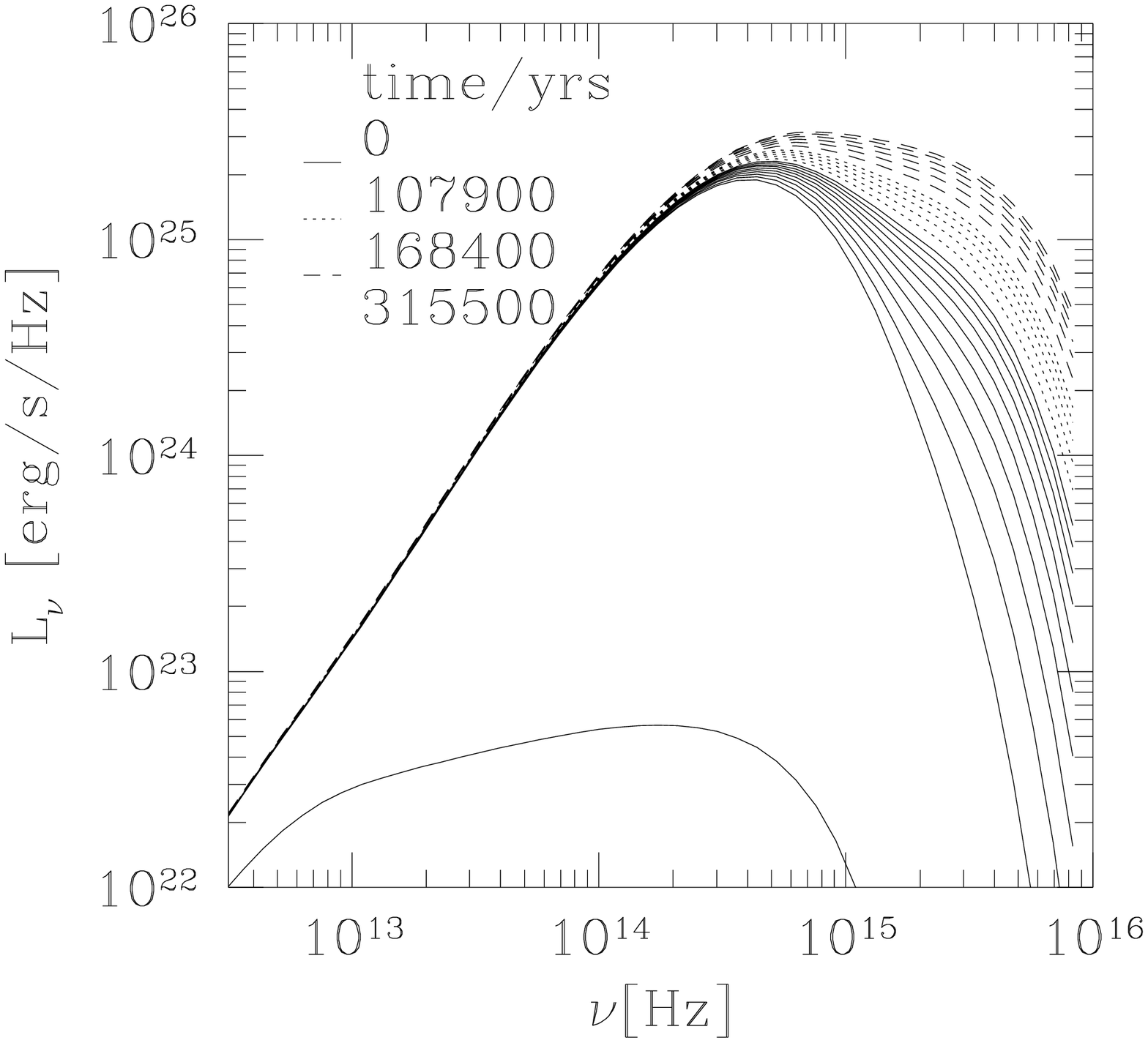}

\placefigure{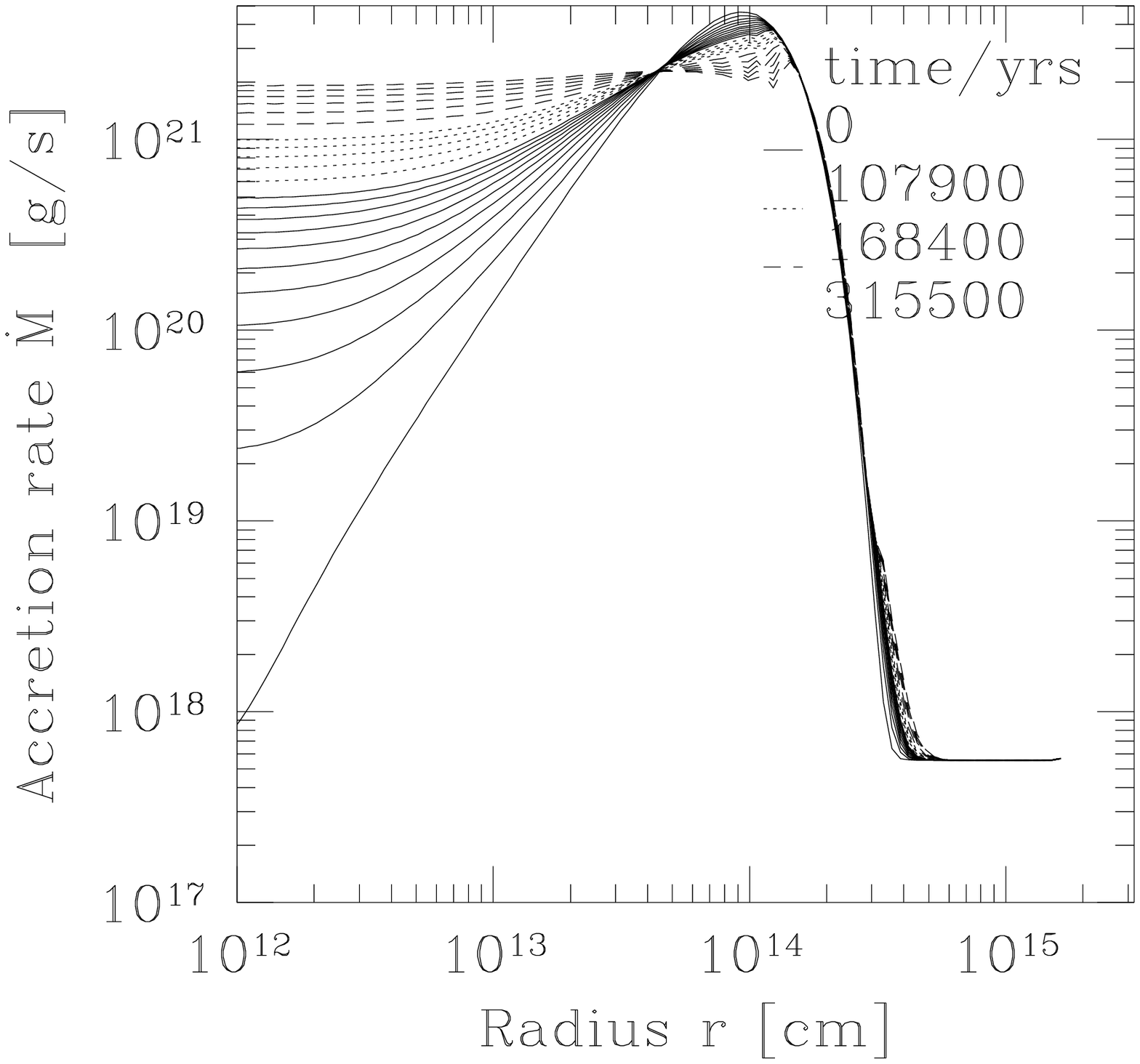}

\section{Preliminary Application To Sgr~A*}

\subsection{The basic scenario}
As our first (and necessarily simplified) application of the results
in the previous section, we now consider the situation in the Galactic
Center.  There exists ample evidence for the presence of a strong wind
in the region around Sgr~A*, with a characteristic velocity $v_w\sim
500-700$ km s$^{-1}$ and mass loss rate $\dot M_w\sim
10^{-3}\;M_\odot$ yr$^{-1}$ (see Melia 1994 and the references cited
therein, particularly Hall, Kleinmann \& Scoville 1982, Geballe et
al. 1991, Krabbe et al. 1991 and Gatley et al. 1986, among others).
In addition, the latest stellar kinematic studies point to a likely
mass of $M\approx 1.8\; (\pm 0.5)\times 10^6\;M_\odot$ for the central
object (Haller et al. 1995), and so it is expected that Sgr~A* is
accreting at a rate of about $10^{(-5)\;{\hbox{\rm to}}\;(-4)}
M_{\sun}/$yr (Ruffert \& Melia 1994) via Bondi-Hoyle accretion.

However, the luminosity of Sgr~A* is tightly constrained to a value of
a few $10^5 L_{\sun}$ (Falcke et al. 1993a; Zylka et al. 1995).  Thus,
if the accretion flow eventually turns into a disk at smaller radii,
and assuming a conversion efficiency of about $10\%$ rest mass energy 
into radiation, we infer a disk accretion rate of a few $10^{-7}M_\odot/$yr.  
If the winds surrounding the central black hole are relatively
uniform, the fluctuations in the accreted specific angular momentum are
sufficiently small that only a small (less than about $5-10$
Schwarzschild radii in size) disk is then expected to form (Melia 1994;
Ruffert \& Melia 1994).  Such a small disk could be consistent with
the low luminosity observed from this region, but it is likely that
gradients in the ambient medium produce a bigger specific angular
momentum in the accreted material, and hence a larger circularization
radius.  This could be due in part to the fact that the Galactic Center 
wind is comprised of several stellar components, which would make
the flow around Sgr~A* non-uniform.  Departures from uniformity increase the
accreted specific angular momentum since the cancellations in the
post bow-shock flow are not complete --- one side tends to dominate over
the other, producing a net accreted angular momentum.  The correspondingly
larger circularization radius produces a bigger and presumably brighter 
disk.  Another way of saying this is that the infalling gas from a
non-uniform medium retains a larger Keplerian energy that must be 
dissipated as it drifts inwards toward the black hole.

It could be that the disk, if large, is highly advective (Narayan et al. 
1995).  The role of advection has usually been ignored in $\alpha$-disk 
theory, but several studies (e.g., by Abramowitz et al. 1988, 1995, and
more recently by Narayan \& Yi 1994) have shown that under some 
circumstances, the internal disk energy can constitute a substantial 
radial flux when it is transported by the inward drift.  As much as 
$90\%$ or more of the locally dissipated gravitational energy flux can 
be carried away in this manner, thereby decreasing the radiated 
luminosity for a given mass accretion rate by as much as an order of 
magnitude, and possibly more.  However, without detailed 
radiative-hydrodynamical simulations, this scenario is at best still
only a possibility.

It has also been suggested that Sgr~A* might be surrounded by a
large, slowly accreting, fossil disk, perhaps formed from the
remnants of a tidally-disrupted star that ventured too close to
the black hole.  In this case one expects that the large-scale
Bondi-Hoyle inflow must then be captured and incorporated into 
the disk.  The latter might remain faint if most of the quasi-spherical
infall circularizes at large radii, for the incorporated gas would
require a fairly long viscous time scale in order to reach the
central black hole and produce the correspondingly high luminosity output 
(Falcke \& Heinrich 1994).  A fossil disk could also act as a 
resevoir for the infalling plasma, thereby delaying the dissipation 
of its energy.

\subsection{Model vs. observation}
Many of the calculations reported in this paper correspond to a
situation similar to that of Sgr~A*.  In order for a particular model
to be relevant to this source, it should be able to account for the
high wind-accretion rate and yet a low luminosity of a few $10^5
L_{\odot}$.  Recent NIR observations of Sgr~A* appear to show 5
different sources within the Sgr~A* error box (Eckart et al. 1995),
though the source separation in this image is smaller than the
telescope's diffraction limit.  If this separation of the infrared
emission from this region is real, the limit on Sgr~A*'s luminosity
could be as low as $L_\nu<3\cdot 10^{21}$ erg/sec/Hz (for an
interstellar absorption of $A_K=3.7$) in the K-band ($\sim10^{14}$
Hz). The upper limit for Sgr~A* at $12.4\mu$m ($10^{13.4}$ Hz) is
0.1~Jy ($0.9\cdot10^{22}$ erg/sec/Hz) (Gezari et al. 1996). Recently
Stolovy, Hayward, \& Herter (1996) claimed a detection of Sgr A* at
8.7$\mu$m (100 mJy dereddened) which may be interesting with respect
to our work but needs further confirmation.  A starving accretion disk
with $\dot M<10^{-8}M_{\sun}$/yr, as used in our simulations, could be
consistent with those numbers when the following additional factors
are taken into account. Requiring that the disk is seen edge-on $(\cos
\theta \la 0.1)$, would reduce the flux displayed in the figures by
another factor 5.  Moreover, there are indications that Sgr~A* is
intrinsically absorbed (Predehl \& Tr\"umper 1994) --- in addition to
the interstellar attenuation --- which would reduce the observed
luminosity further.

A second important factor to take into account when comparing particular
models to the observations is the time scale required for a change in the
radial accretion rate.  If the main components in the Bondi-Hoyle 
accretion are the stellar winds of central He I stars, the accreting
medium should have been there a certain fraction of the stars' 
lifetime ($\sim 3-7\times 10^6$ years, Krabbe et al. 1995), though probably
not longer than this. We would therefore expect that a typical
time scale for the accretion process should be longer than of order
a million years if a fossil disk is involved.

The luminosity and time scale constraints would therefore seem to
rule out situations in which the wind is captured by the disk at
small radii, because they produce spectra that are too luminous in
the NIR and IR very quickly after the onset of Bondi-Hoyle accretion
(Figs. \ref{fig-narrow.xi=1} \& \ref{fig-narrow.xi=0}). In other words,
the post-bow shock flow cannot be carrying a small specific angular
momentum (which would lead to circularization at small radii), but
rather should have sufficient azimuthal velocity to allow it to merge
with the fossil disk near its outer edge.  However, models
with a low angular momentum wind that is captured far out in the disk
fare no better (as far as Sgr~A* is concerned;  see Fig. \ref{fig-wide.xi=0}) 
because they violate the IR limits due to
their instant transformation of kinetic energy into heat and thence
into IR radiation.

It appears from our analysis of the calculations that the only class
of models that might be consistent with the observations of Sgr~A*
are those in which the angular momentum of the wind is large enough 
that the infall circularizes at a radius $\ga10^{16}$ cm and settles 
gently onto the disk (Figs. \ref{fig-wide.xi=1} \& \ref{fig-hurricane}). 
In addition, the $\alpha$-parameter must be small ($\alpha\ll1$) in 
order to prolong the transfer of the deposited wind material into the
center. The `hurricane' model with $\alpha\la10^{-4}$ seems to be 
a viable scenario, though this model does not even
require the presence of a fossil accretion disk to stop the infall
because the high angular momentum wind will circularize and stop at
larger radii anyway.  If it exists, a starving, fossil disk may even
lie closer to the black hole and not be coupled directly
with the large scale disk formed by the wind infall.

\section{Discussion}
In this paper we have rederived the basic equations for an 
$\alpha$-accretion disk with the consideration of wind infall. Besides a source
term for the mass, the wind infall also introduces a dynamical viscosity
$\nu_{\rm w}$ to the basic equations when its angular momentum is
sub-Keplerian. We have calculated the evolution of an accretion disk
with wind infall for various sets of parameters, leading to some
interesting effects not present in conventional accretion disk
theory. The situation we considered in detail is that of a fossil
accretion disk, which intercepts a wind with a much higher accretion
rate than the former.  As long as the viscosity parameter in such a disk
is very small ($\alpha\ll 1$), the time during which the surface 
density of the disk changes is much longer than the dynamical time scale
because of its large mass storage capability.  In this case, 
the disk is not blown away and it can incorporate winds with very high
mass accretion rates. A large, non-coaxial angular momentum in the wind
may, however, lead to a warping of the disk which we ignore here.

The overall evolution of the disk may be described in terms of two basic
classical solutions: (a) that of an almost empty disk that is fed from the
outside and (b) an initial ring of matter that evolves with time and
where most of the matter is flowing in, while a small fraction flows
outwards carrying away the angular momentum.
If the intercepted wind has a low specific angular momentum, the disk will
shrink in the radial direction on a short time scale determined by 
$\Sigma/\dsigw$ and de-couple from its outer region, forming a
gap. The matter piles up until viscous processes in the disk become
important and the inner regions are filled up like in the empty disk
evolution. For a high specific angular momentum wind, the disk
will also have a radial outflow in its outer portion 
like in the ring evolution.  The interaction between wind and disk leads to an
additional heating of the disk on its surface and a moderate increase
in the viscosity and therefore a faster evolution; the
effect is stronger for a sub-Keplerian wind.

The most important changes with respect to the conventional solutions
concern the emitted spectrum: the emission at lower-frequencies is 
enhanced. This may produce either a flattening of the spectrum or
even a strong low-frequency bump (in AGN the bump would be produced in
the IR). This bump can be as luminous as the primary bump produced in
the inner region of the disk or even brighter. The most drastic effect
is due to the angular momentum deposited by the wind. When the wind has
a Keplerian velocity and settles gently onto the disk ($u_{\rm z}=0$), the
formation of the low-frequency bump may take a long time
$\Sigma/\dsigw$, while a wind with a low specific angular momentum and 
a high $u_{\rm z}$ will release its kinetic energy very quickly 
and the low-frequency bump will appear almost
instantaneously. In both cases, however, it will take a viscous time
$R^2/\nu$ for the increased accretion rate in the outer part
to reach the inner region and lead to an increase of the
(usually dominating) high-frequency emission.

The process described here has many potential applications. The
production of an IR bump by an infalling wind, perhaps from the
nuclear star cluster or a circumnuclear starburst, would be an
interesting --- although as yet undeveloped --- signature to
investigate.  A scenario such as this can also be invoked to account
for some observational characteristics in winds onto protostellar
disks.  In this paper, we have briefly discussed the application of
our model to the Galactic Center source Sgr~A*.

The presence of strong stellar winds and a high central mass
concentration in the Galactic Center appear to be inconsistent with
the low Infrared luminosity of the putative supermassive black hole,
Sgr~A*.  We have considered the possibility that this object may be
surrounded by a fossil disk accreting at a rate of
$10^{-8}M_{\sun}/$yr, while it is intercepting the incoming wind with
$\dot M_{\rm w}\la10^{-4}M_{\sun}$/yr.  Current observational limits
for Sgr~A* are very restrictive and we have found that they are not
consistent with most of the wind specific angular momentum
configurations we have considered here.  As the estimates for the
bolometric luminosity of Sgr~A* may change, the direct observations in
the IR and NIR are the most critical limits to be observed in this
respect and we find that any fossil disk would become too luminous too
quickly in those bands. We can only reconcile the data with the models
if the wind circularizes on a scale of $\ga10^{16}$ cm and if the
$\alpha$-parameter within the disk is $\la10^{-4}$.  Without detailed
numerical simulations, these numbers are at best only estimates.  Even
so, it seems safe to conclude that if a fossil disk is present around
Sgr~A*, then it must have a very low $\alpha$ and the wind must carry
a large specific angular momentum in order to describe Sgr~A*
self-consistently.  This rules out a large region of the phase space
available for possible models of Sgr~A*.

In a future extension of this work, we plan to couple our calculations 
to 3D hydrodynamical simulations for the large scale Bondi-Hoyle
accretion (Paper II) in order to get a better estimate for
the radial distribution of the wind infall. As we have seen here,
the specific angular momentum of the wind appears to be
the most critical parameter.  Further improvements of our model
will include a more accurate description of the opacity and the
possible warping of the disk by a very strong, asymmetric wind.

\acknowledgements This research was supported in part by the NSF
under grant PHY 88-57218, and by NASA under grants NAGW-2518 and NAG8-1027.

\clearpage
\figcaption{\label{fig-empty} \label{fig-empty.s.sig.ps}
\label{fig-empty.l.sig.ps}
Surface density evolution of an almost
empty disk that is fed from the outer boundary with
$10^{-2}M_{\sun}$/yr. The analytical stationary solution is given by
the dashed-dotted line.  The lower panel (the zero $\Sigma$
distribution is outside the plot range here) shows the evolution on an
extended time scale compared to the upper panel.}

\figcaption{\label{fig-gauss}\label{fig-gauss.sig.ps}\label{fig-gauss.mdot.ps}
Surface density and accretion rate evolution of a 
Gaussian ring with no external
feeding. For comparison, two appropriately scaled stationary solutions
are plotted, the upper curve corresponding to $\dot M=10^{-2}\;
M_\odot$/yr and the lower one to an accretion rate roughly 5 
times lower, which is the actual accretion rate in the disk.}

\figcaption{\label{fig-wide.xi=1}\label{fig-wide.xi=1.sig.ps}\label{fig-wide.xi=1.lnu.ps}
Surface density and emitted spectrum evolution of a fossil
accretion disk with strong wind deposition; the wind has Keplerian
azimuthal velocity ($M_\bullet=10^6M_{\sun}$, $\dot M_0 =
10^{-8}M_{\sun}$/yr, $\alpha=10^{-4}$, $\dot M_{\rm
w}=10^{-4}M_{\sun}/$yr, $\xi=1$, and $r_{\rm w}=90000 R_{\rm g}$).}

\figcaption{\label{fig-hurricane}\label{fig-hurricane.sig.ps}\label{fig-hurricane.lnu.ps}
Surface density and emitted spectrum evolution of a fossil
accretion disk with strong wind deposition only within the outer ring
(a ``hurricane''); the wind has Keplerian azimuthal velocity
($M_\bullet=10^6M_{\sun}$, $\dot M_0 = 10^{-8}M_{\sun}$/yr,
$\alpha=10^{-4}$, $\dot M_{\rm w}=10^{-4}M_{\sun}/$yr, $\xi=1$, and
$r_{\rm w}=30000-90000 R_{\rm g}$).}

\figcaption{\label{fig-wide.xi=0}\label{fig-wide.xi=0.sig.ps}\label{fig-wide.xi=0.lnu.ps}
Surface density and emitted spectrum evolution of a fossil
accretion disk with strong wind deposition; the wind has zero
azimuthal velocity ($M_\bullet=10^6M_{\sun}$, $\dot M_0 =
10^{-8}M_{\sun}$/yr, $\alpha=10^{-4}$, $\dot M_{\rm
w}=10^{-4}M_{\sun}/$yr, $\xi=0$, and $r_{\rm w}=90000 R_{\rm g}$).}

\figcaption{\label{fig-narrow.xi=1}
\label{fig-narrow.xi=1.sig.ps}\label{fig-narrow.xi=1.lnu.ps}\label{fig-narrow.xi=1.mdot.ps}
Surface density, emitted spectrum, and accretion rate 
 evolution of a fossil accretion disk with strong wind deposition at
small radii; the wind has a Keplerian azimuthal velocity
($M_\bullet=10^6M_{\sun}$, $\dot M_0 = 10^{-8}M_{\sun}$/yr,
$\alpha=10^{-4}$, $\dot M_{\rm w}=10^{-4}M_{\sun}/$yr, $\xi=1$, and
$r_{\rm w}=900 R_{\rm g}$). For comparison we have plotted  a
stationary solution on top of the $\Sigma$ curves (dashed-dotted line,
top figure).}

\figcaption{\label{fig-narrow.xi=0}\label{fig-narrow.xi=0.sig.ps}
\label{fig-narrow.xi=0.lnu.ps}\label{fig-narrow.xi=0.mdot.ps}
Surface density, emitted spectrum,  and accretion rate 
 evolution of a fossil accretion disk with strong wind deposition at
small radii; the wind has zero angular momentum
($M_\bullet=10^6M_{\sun}$, $\dot M_0 = 10^{-8}M_{\sun}$/yr,
$\alpha=10^{-4}$, $\dot M_{\rm w}=10^{-4}M_{\sun}/$yr, $\xi=0$, and
$r_{\rm w}=900 R_{\rm g}$). For comparison we have plotted a
stationary solution on top of the $\Sigma$ curves (dashed-dotted line,
top figure). The height in the `gap' is an artifact, as there is
no disk anymore in this region.}

\clearpage
\onecolumn
\begin{figure}
\epsscale{0.5}
\plotone{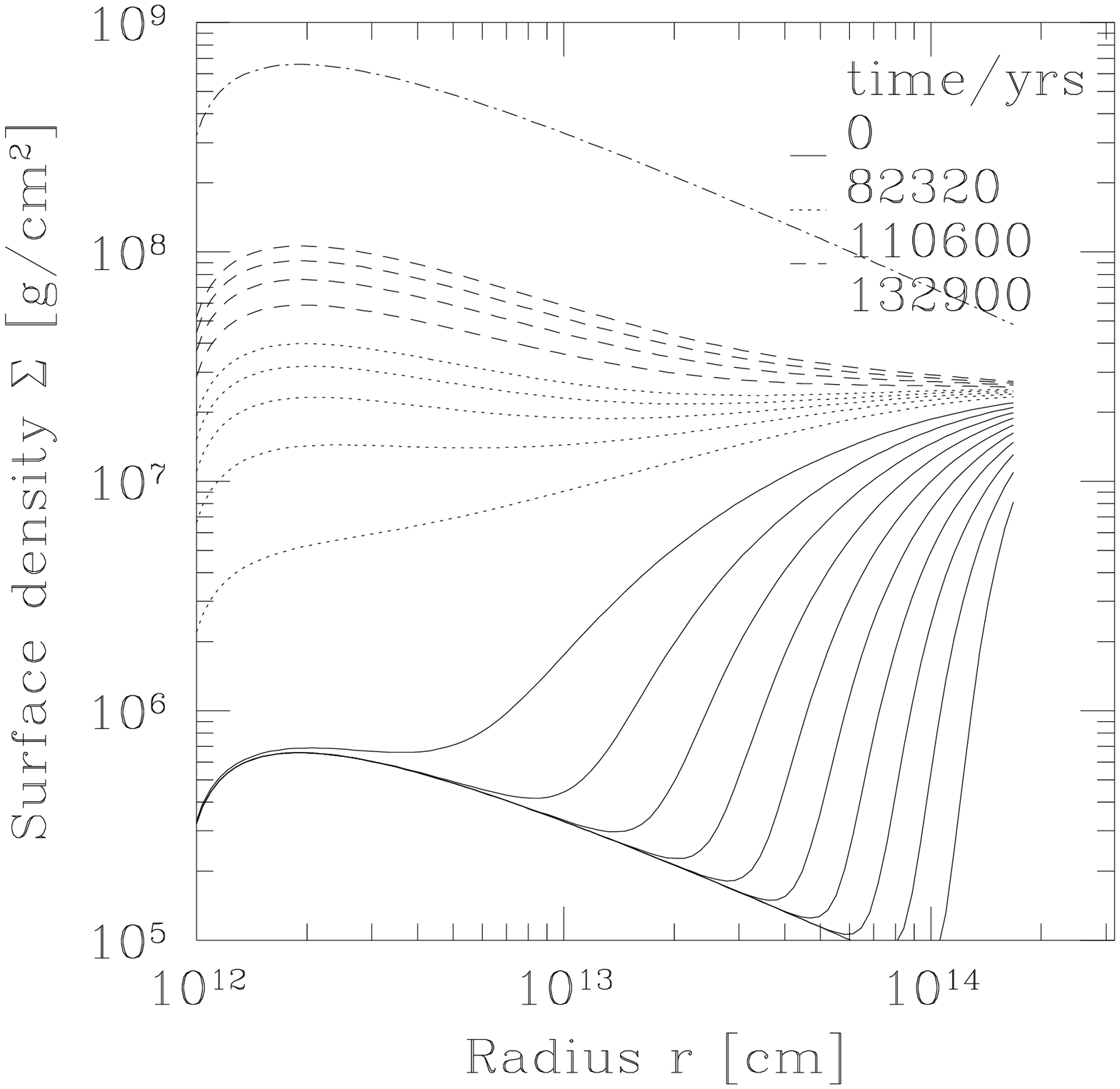}

\epsscale{0.5}
\plotone{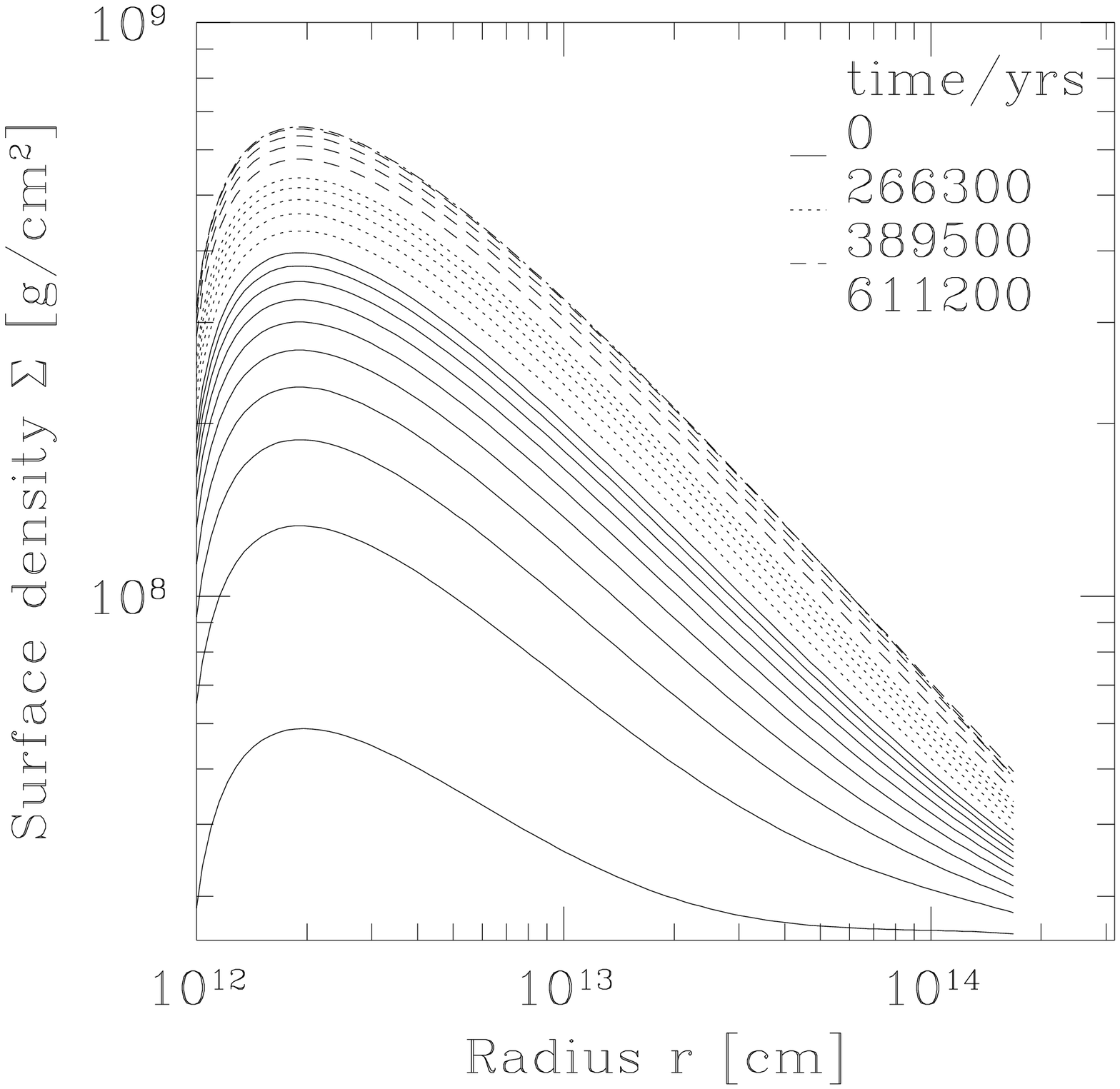}

\centerline{Figure \ref{fig-empty.l.sig.ps}}
\end{figure}

\clearpage

\begin{figure}
\epsscale{0.5}
\plotone{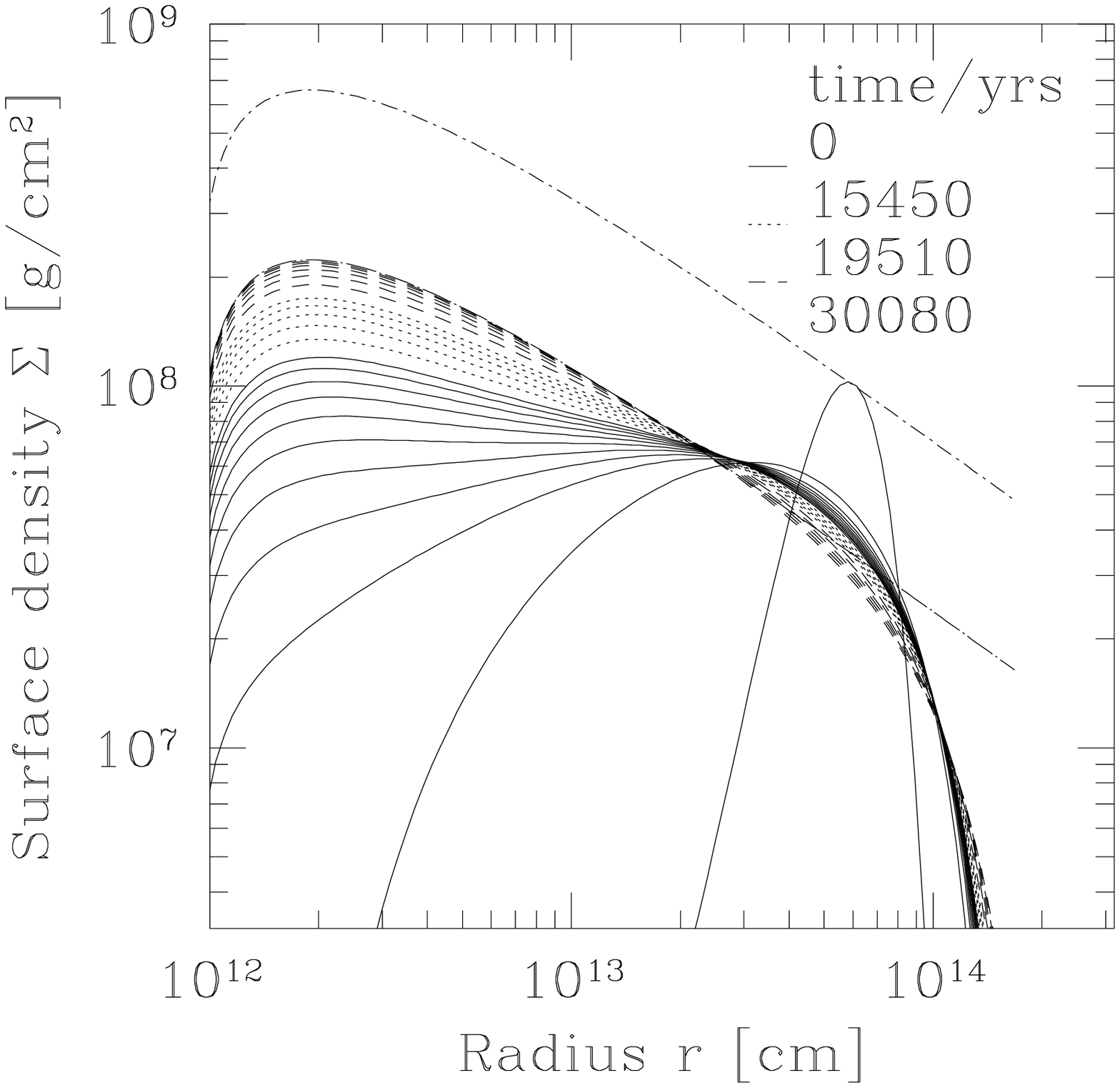}

\epsscale{0.5}
\plotone{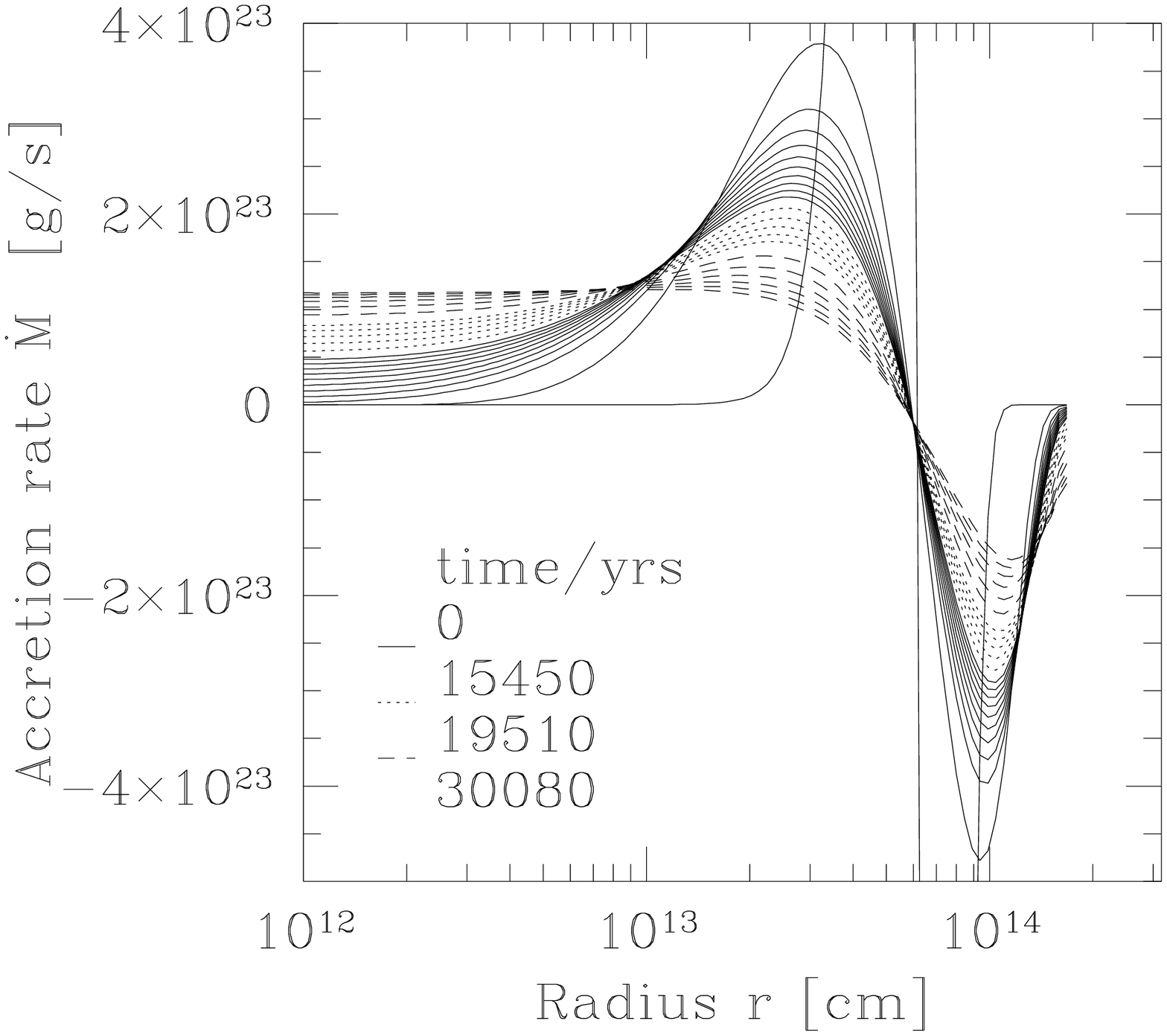}

\centerline{Figure \ref{fig-gauss.sig.ps}}
\end{figure}

\clearpage

\begin{figure}
\epsscale{0.5}
\plotone{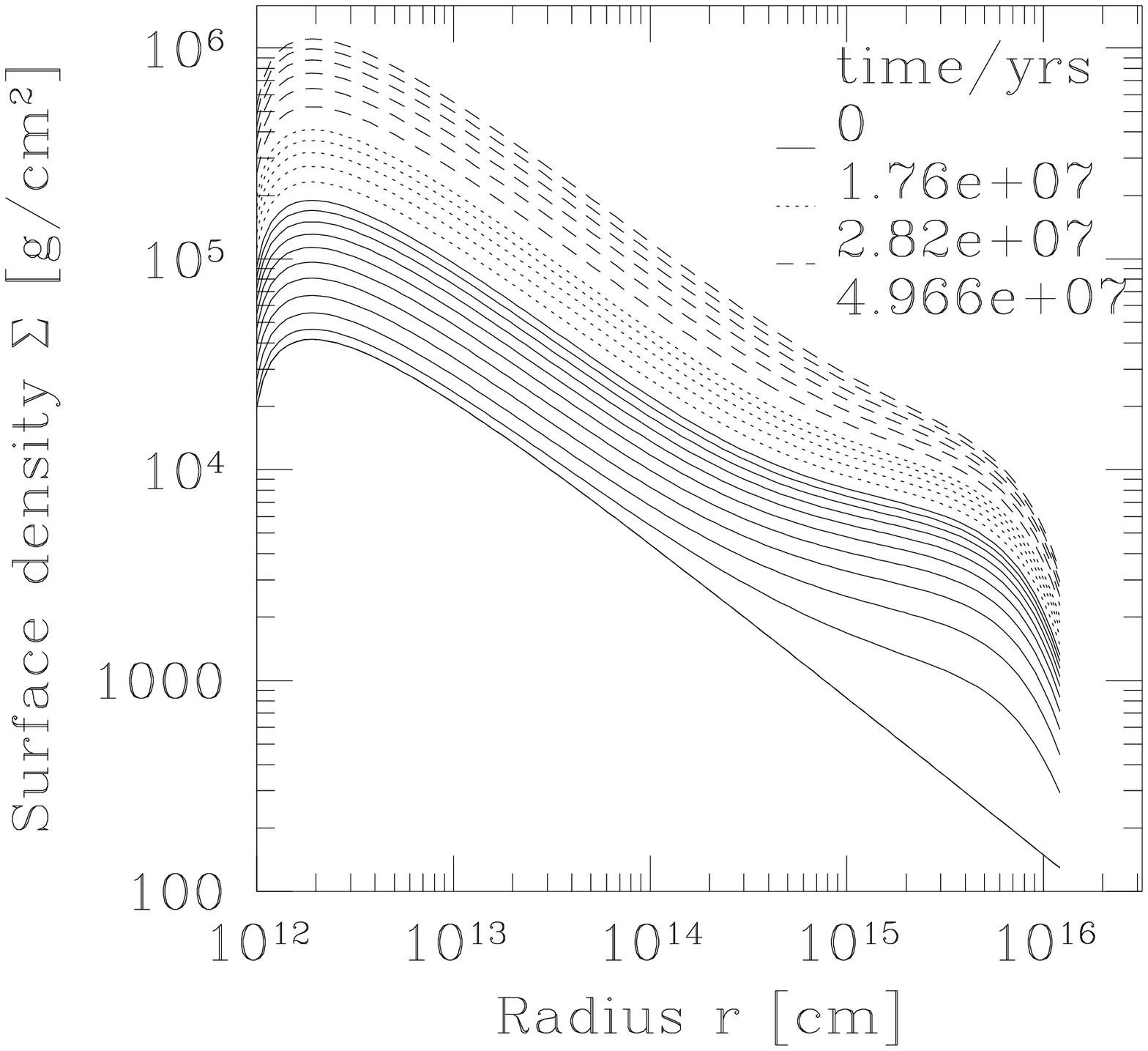}

\epsscale{0.5}
\plotone{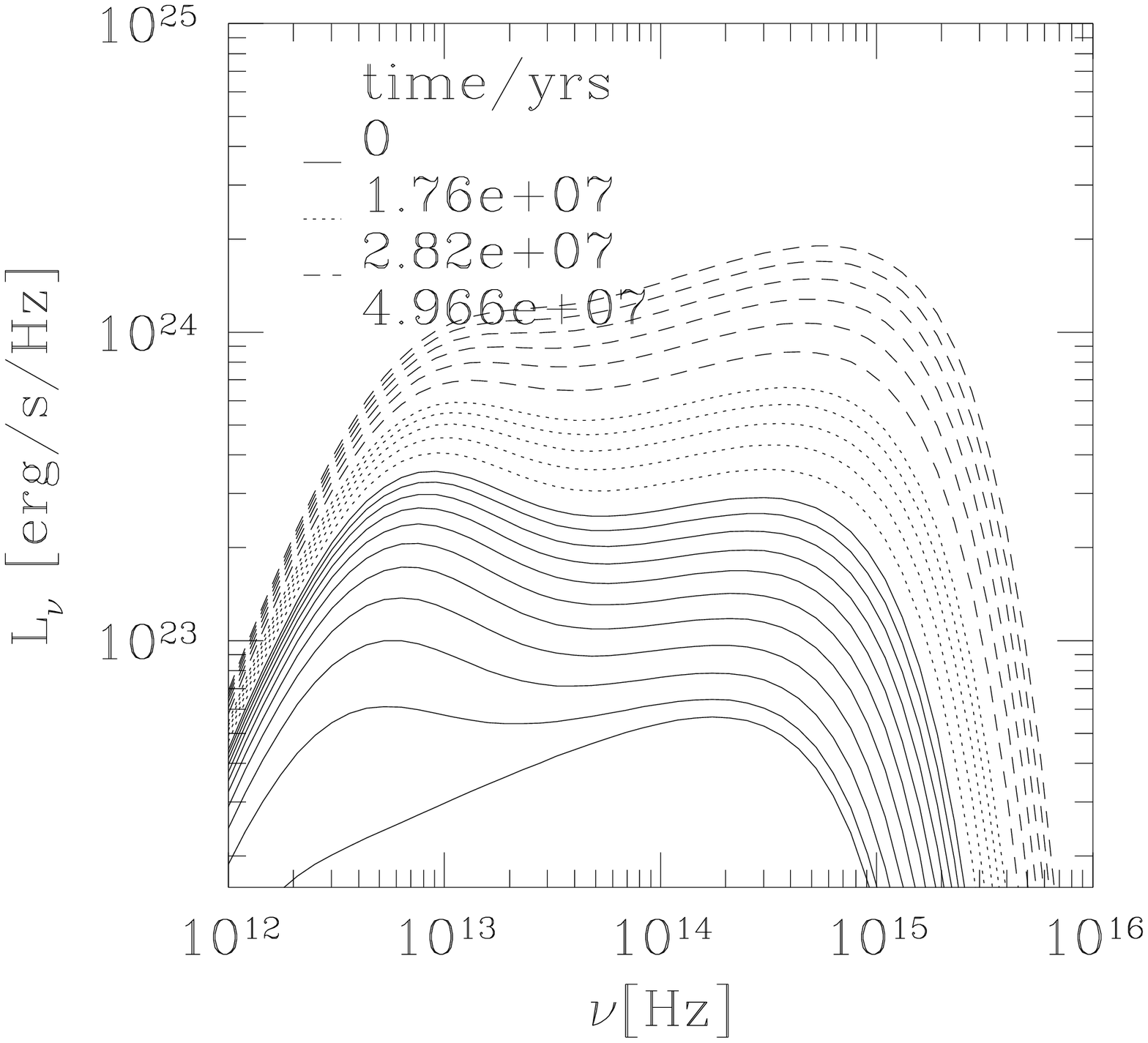}

\centerline{Figure \ref{fig-wide.xi=1.sig.ps}}
\end{figure}

\clearpage

\begin{figure}
\epsscale{0.5}
\plotone{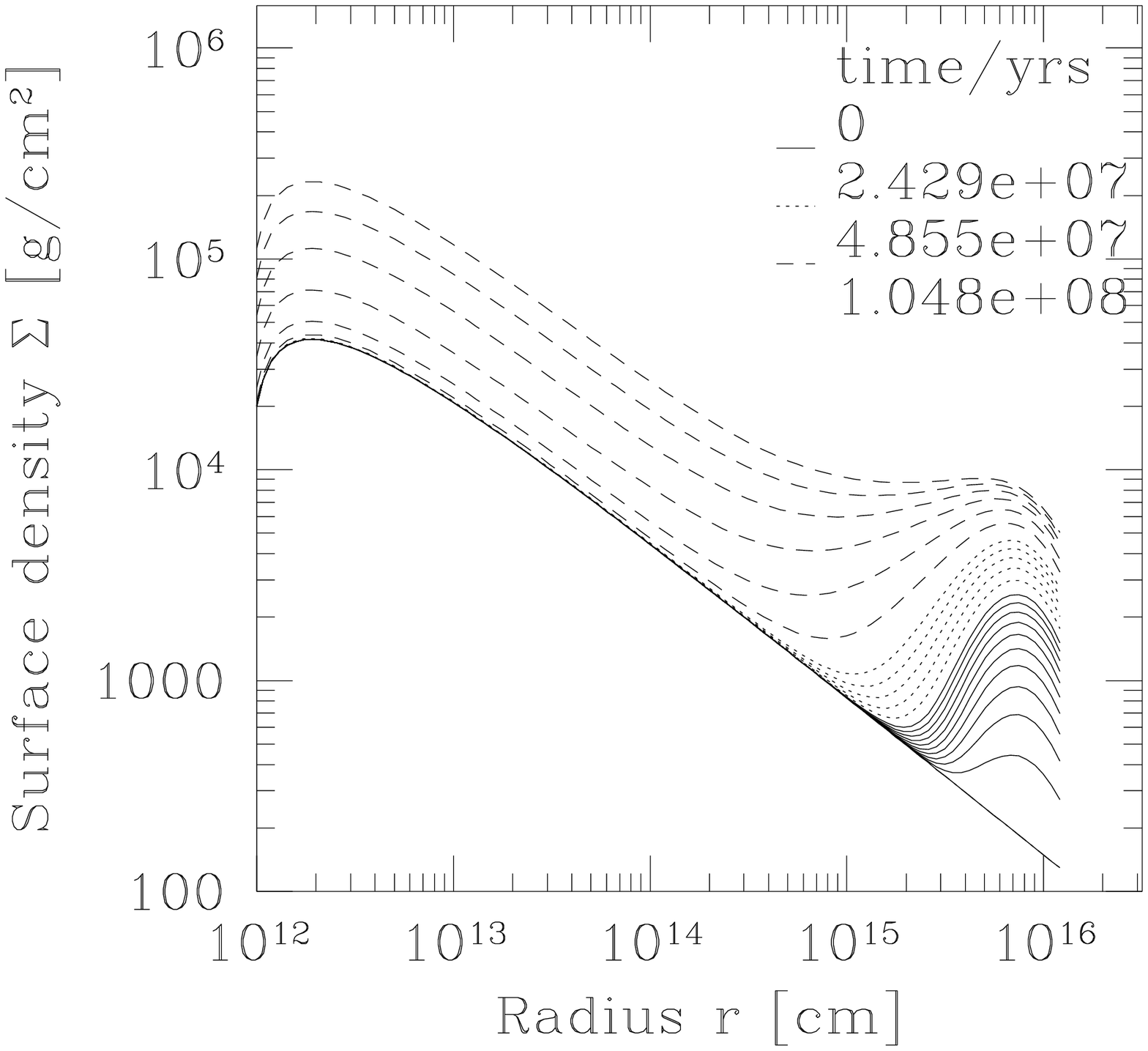}

\epsscale{0.5}
\plotone{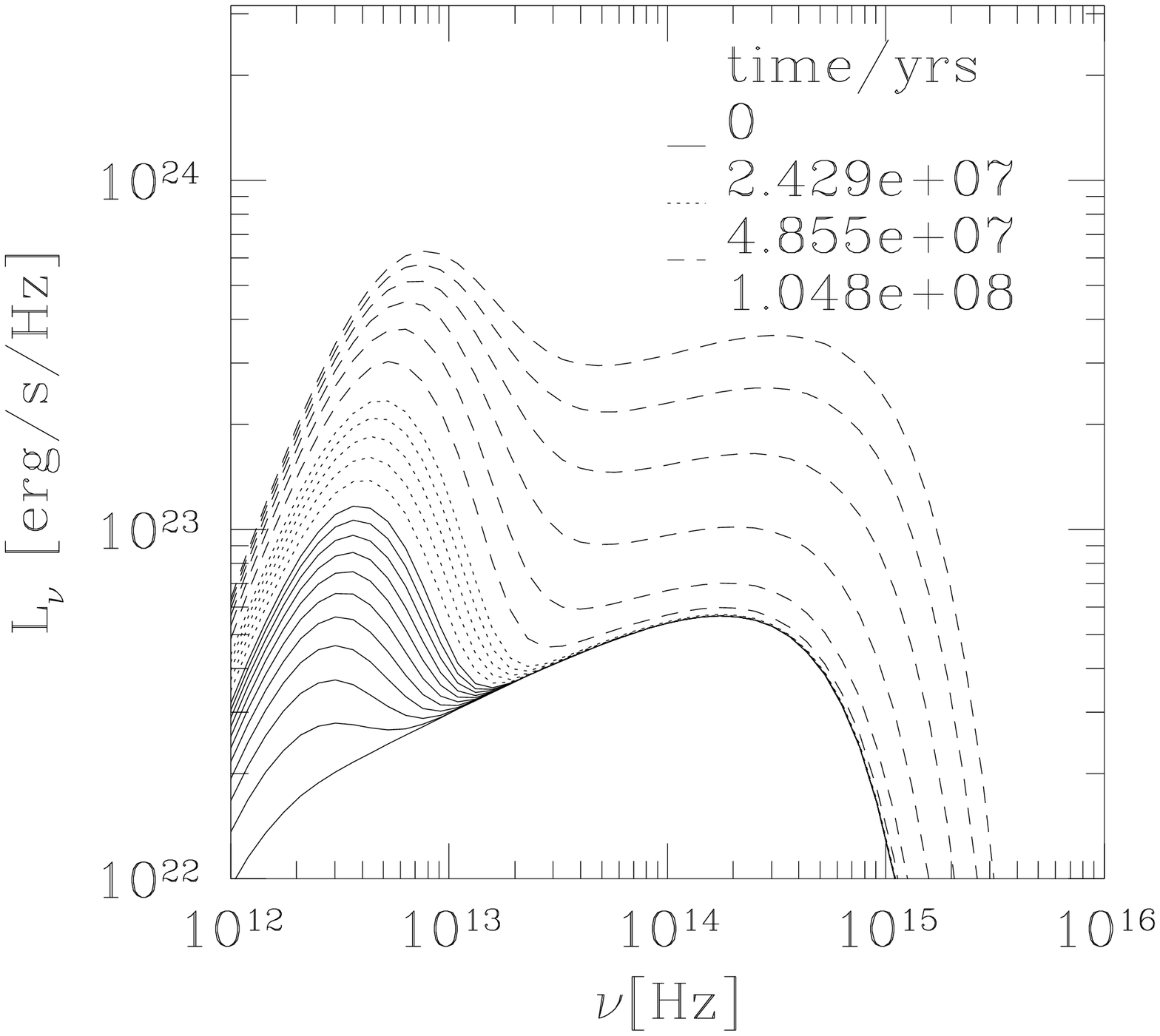}

\centerline{Figure \ref{fig-hurricane.sig.ps}}
\end{figure}

\clearpage

\begin{figure}
\epsscale{0.5}
\plotone{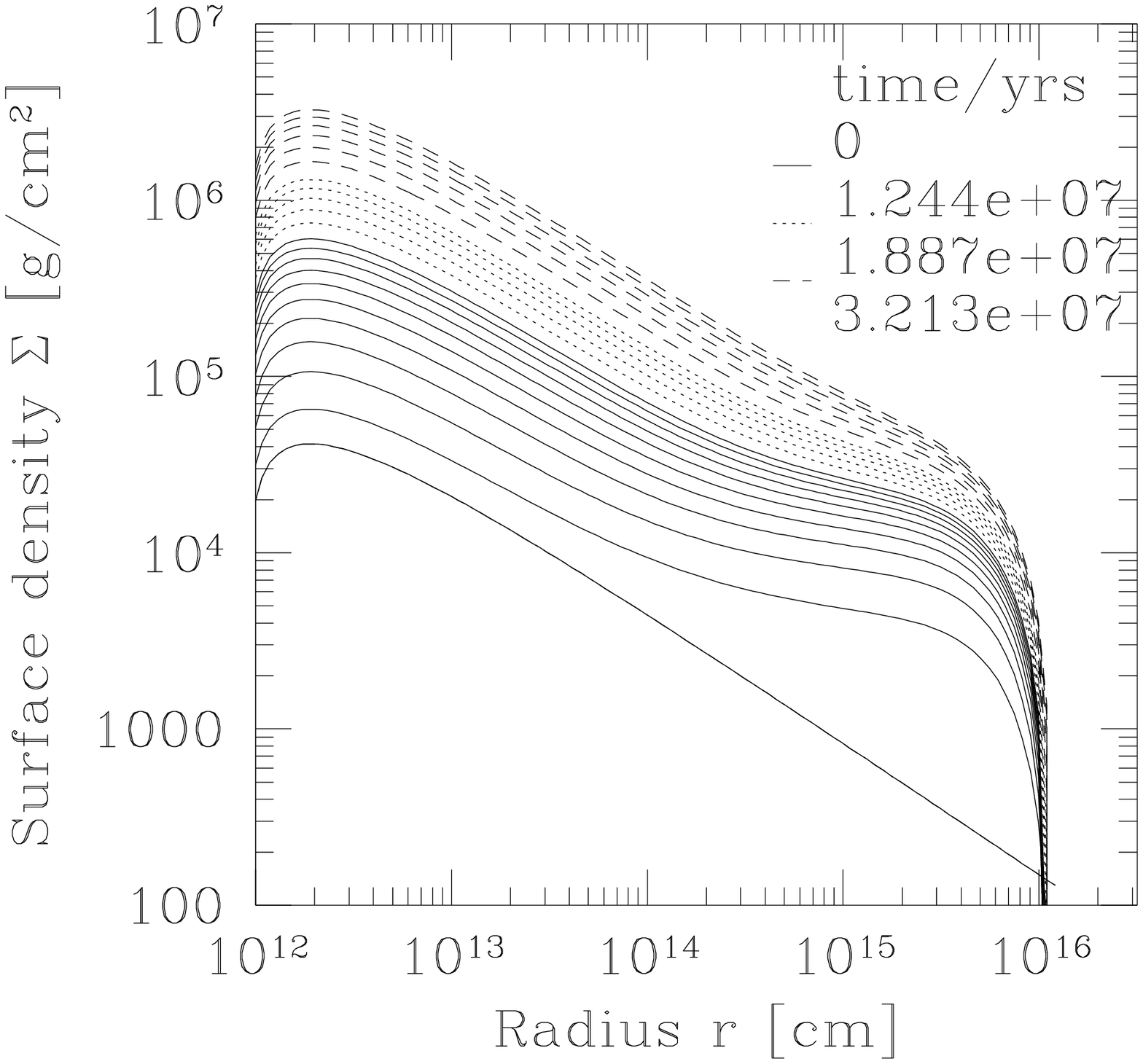}

\epsscale{0.5}
\plotone{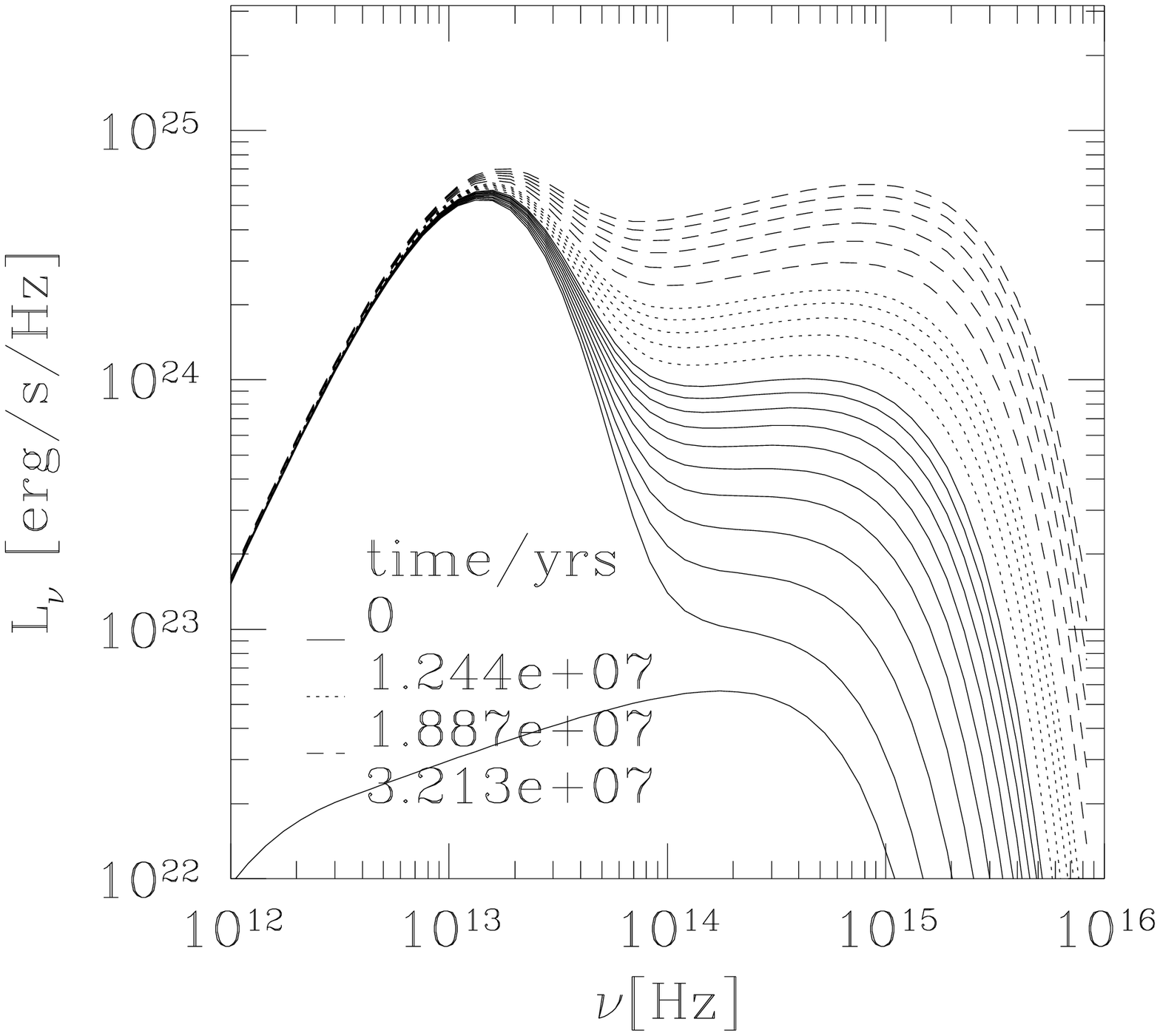}

\centerline{Figure \ref{fig-wide.xi=0.sig.ps}}
\end{figure}

\clearpage

\begin{figure}
\epsscale{0.4}
\plotone{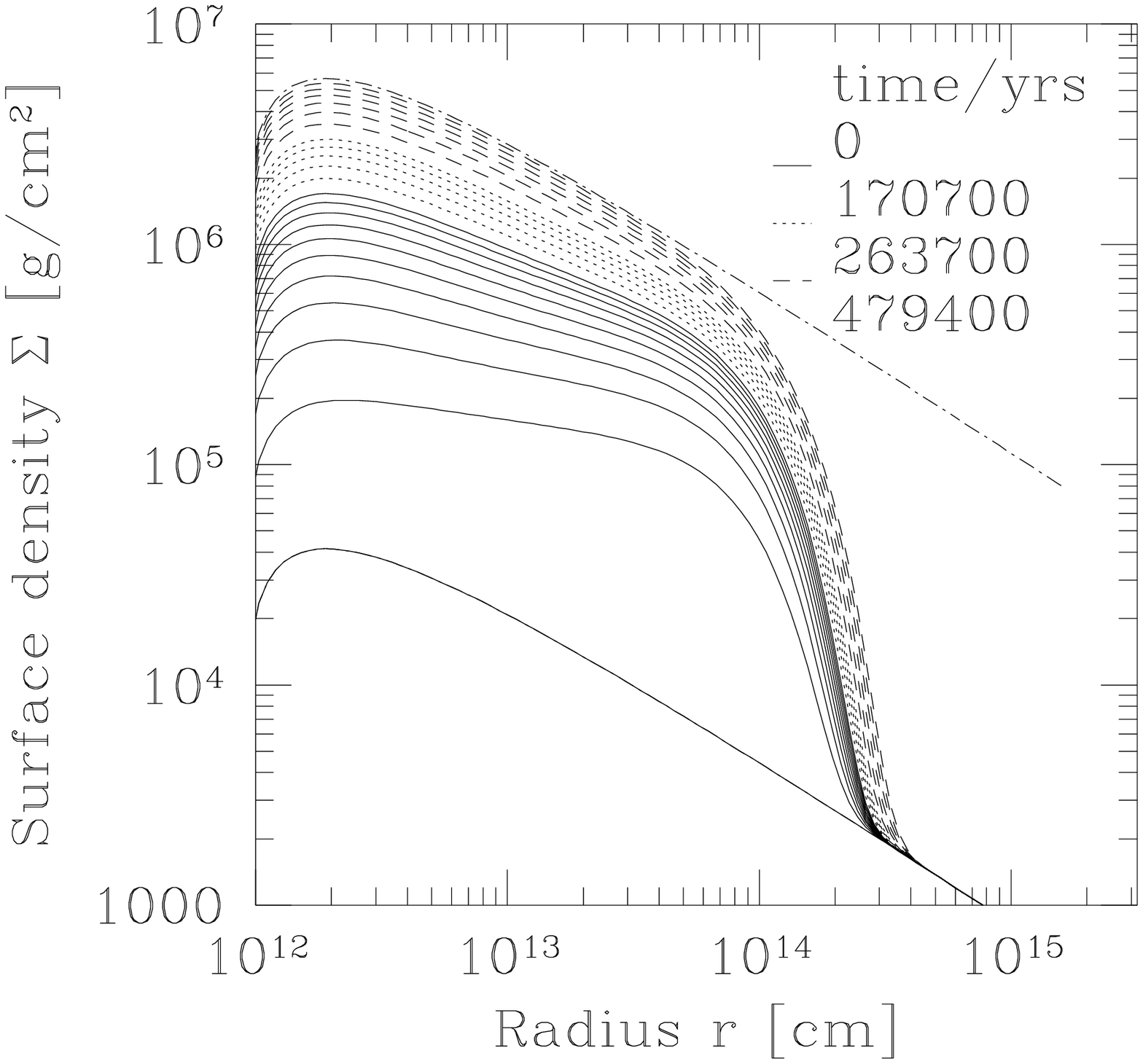}

\epsscale{0.4}
\plotone{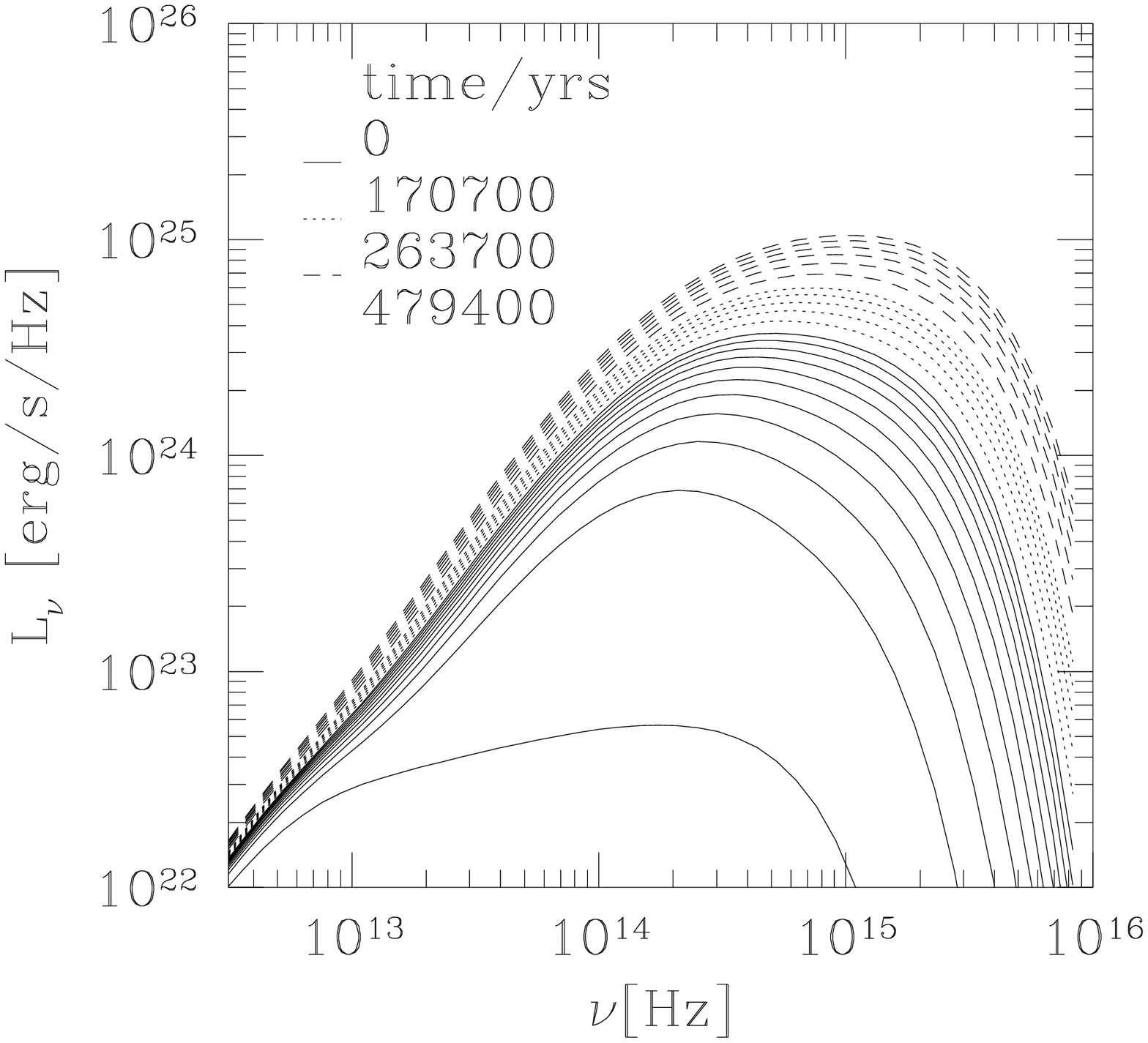}

\epsscale{0.4}
\plotone{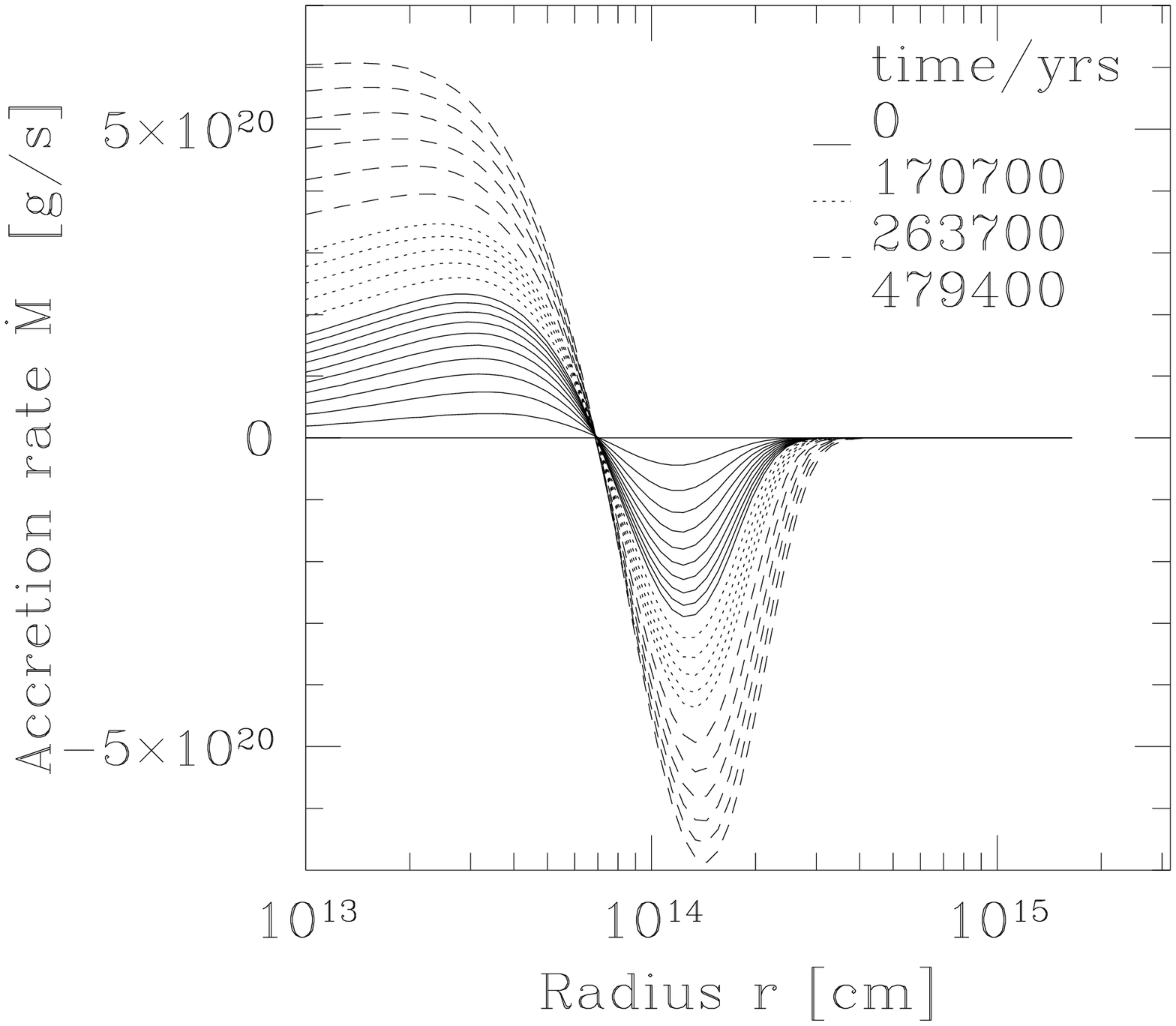}

\centerline{Figure \ref{fig-narrow.xi=1.sig.ps}}
\end{figure}

\clearpage

\begin{figure}
\epsscale{0.4}
\plotone{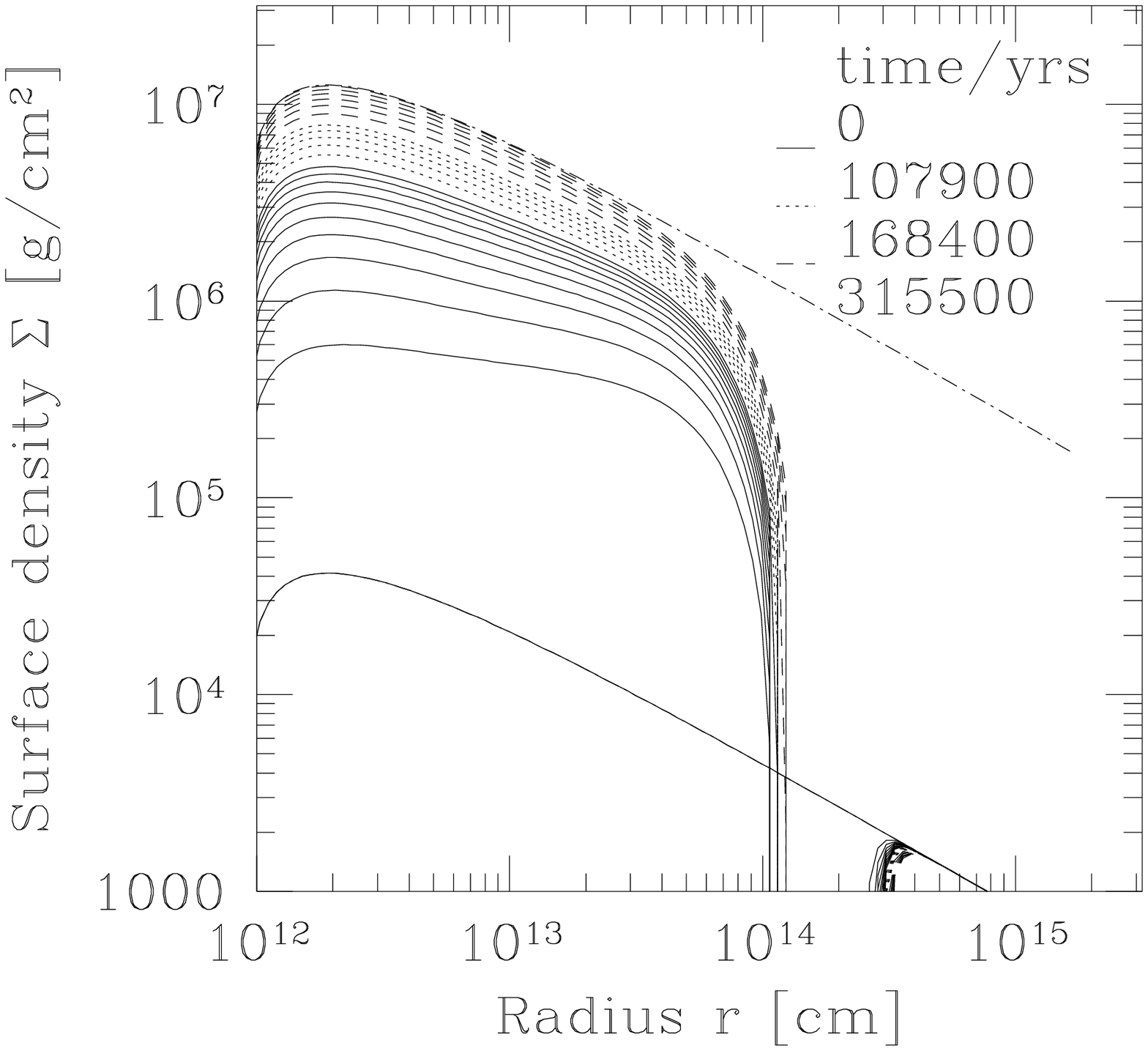}

\epsscale{0.4}
\plotone{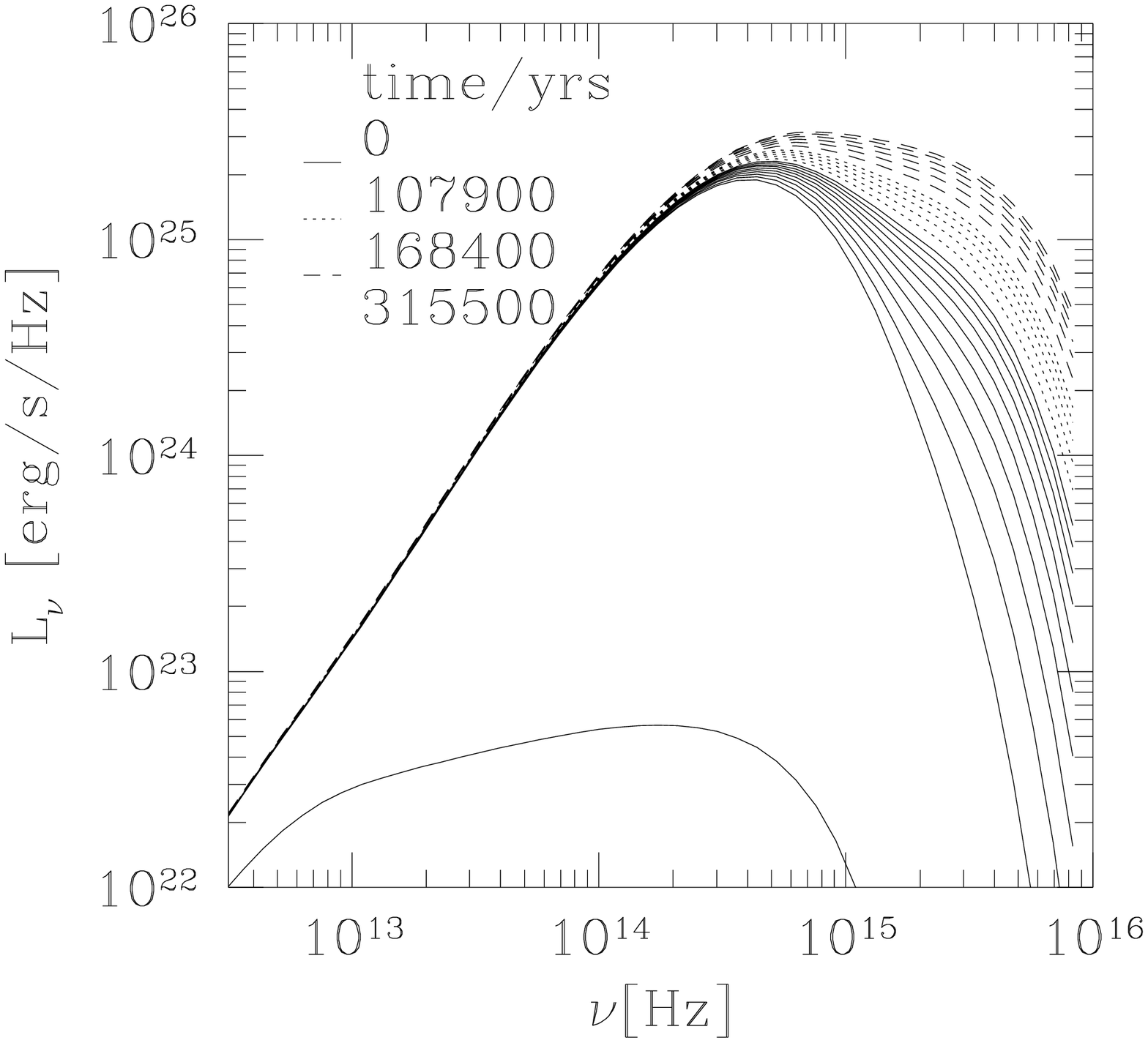}

\epsscale{0.4}
\plotone{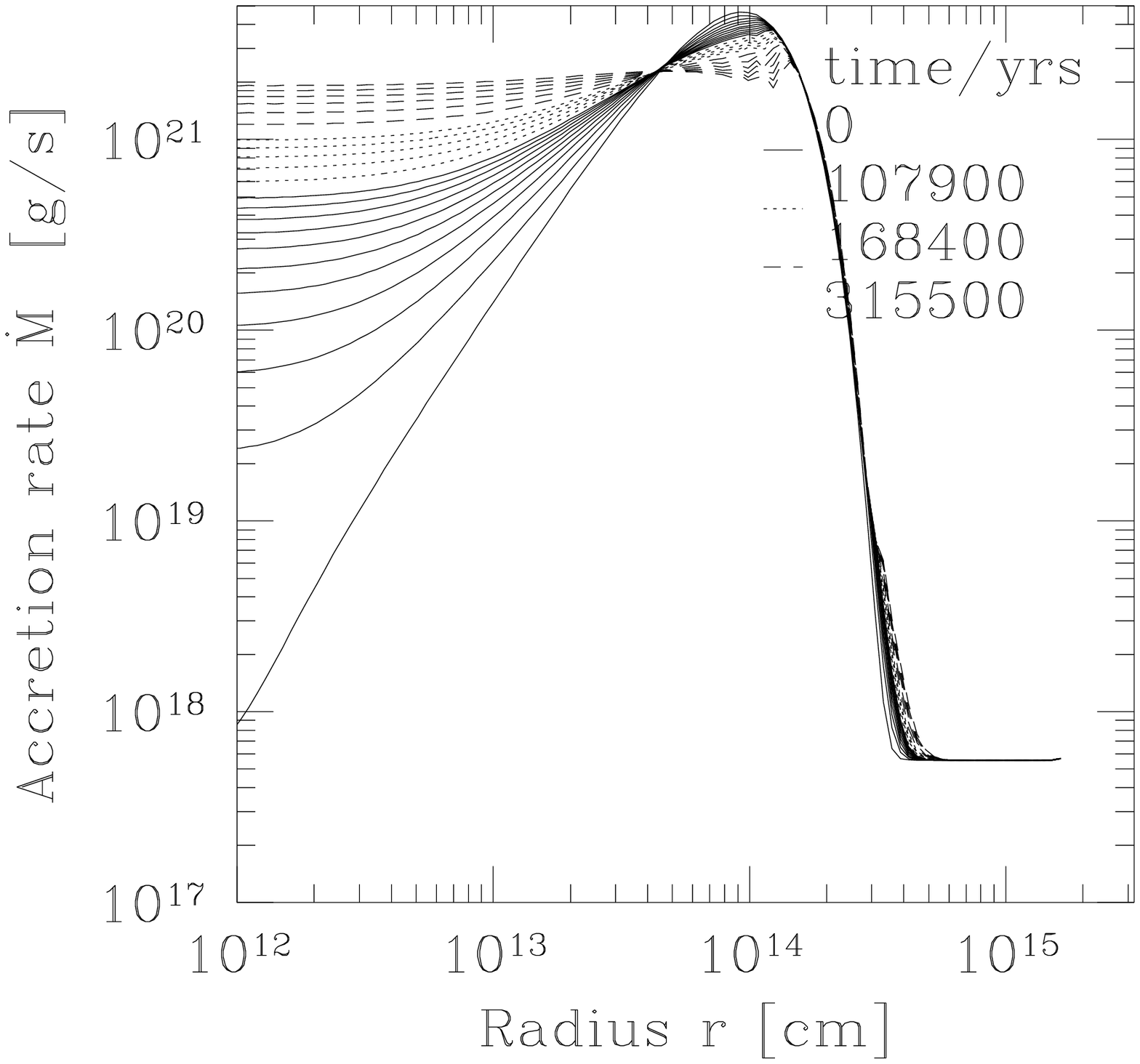}

\centerline{Figure \ref{fig-narrow.xi=0.sig.ps}}
\end{figure}

\end{document}